\documentclass[preprint,nofootinbib,superscriptaddress,amsmath,amssymb]{revtex4}

\usepackage{amsmath} 
\usepackage{verbatim} 
\usepackage{graphicx} 
\usepackage{feynmp}	
\usepackage{subfigure}

\begin{document}

\begin{fmffile}{graphs}
\title{General Covariance Constraints on Cosmological Correlators}

\author{Cristian Armendariz-Picon}
\affiliation{Department of Physics, Syracuse University, Syracuse, NY 13244-1130, USA}

\author{Jayanth T. Neelakanta}
\affiliation{Department of Physics, Syracuse University, Syracuse, NY 13244-1130, USA}

\author{Riccardo Penco}
\affiliation{Physics Department and Institute for Strings, Cosmology, and Astroparticle Physics, 
Columbia University, New York, NY 10027, USA}

\begin{abstract}
We study the extent to which diffeomorphism invariance restricts the properties of the primordial perturbations in single scalar field  models. We derive a set of  identities that constrain the connected correlators of the cosmological perturbations, as well as the one-particle-irreducible vertices of the theory in any gauge. These identities are the analogues of Slavnov-Taylor identities in gauge theories, and follow  essentially from  diffeomorphism invariance alone. Yet because quantization requires diffeomorphism invariance  to be broken, they not only reflect invariance under diffeomorphisms, but also how the latter has been broken by gauge fixing terms. In order to not lose the symmetry altogether, we cannot simply set some fields to zero, as is usually done in cosmological perturbation theory, but need to decouple them smoothly and make sure that they do not contribute to cosmological correlators in the decoupling limit.   We use these identities to derive a set of consistency relations between bispectra and power spectra of cosmological perturbations in different gauges. Without additional assumptions, these consistency relations just seem to reflect the redundancy implied by diffeomorphisms. But when combined with analyticity, in a formulation of the theory in which auxiliary fields have been integrated out, we recover novel and previously derived relations that follow from invariance under both time and spatial diffeomorphisms.
 
\end{abstract}
\maketitle

\section{Introduction}

Whether local symmetries such as diffeomorphism invariance have any physical content has been a subject of intense debate ever since the inception of general relativity.  Indeed, already in 1917, Erich Kretschmann argued  that the principle of general covariance is physically vacuous: Any non-covariant theory ought to be made covariant without changing any of its physical predictions \cite{Kretschmann:1917}, and,  conversely, any covariant theory can be made non-covariant by gauge fixing, a process  that preserves the physical implications of the covariant theory.  

Yet diffeomorphism  (or gauge) invariance does seem to have significant physical implications. In general relativity, for instance, diffeomorphism invariance enforces the equivalence principle \cite{Brout:1966oea,DeWitt:1967uc}. More generally, the invariance of gauge theories under gauge transformations severely constrains the structure of the counter-terms, and plays a crucial role in the demonstration that these theories are renormalizable. Indeed, local symmetries such as diffeomorphism  and gauge invariance are the basis of our understanding of all interactions,  both in the standard model and general relativity. But whether any of the physical implications of these theories do really follow from gauge invariance, or whether they are a consequence of the field content and the residual global symmetries that gauge fixing allows us to preserve, often remains somewhat obscure. 

In this article we study the extent to which diffeomorphism invariance  constrains the properties of the primordial perturbations.  We formulate a set of identities that relate different connected correlators, and also different  one-particle-irreducible (1PI) diagrams in general relativity coupled to a scalar field. These identities belong to a family of relations connected to symmetry, and  have appeared under different names in different contexts.  They are known as  Slavnov-Taylor, Ward-Takahashi or Dyson-Schwinger equations, although they are all basically equivalent: They all express the invariance of the theory under diffeomorphisms.    

These identities have to be interpreted appropriately, however. In order to quantize a theory with a local symmetry, such as diffeomorphism invariance in the case at hand, the symmetry has to be explicitly broken by gauge-fixing terms. Hence,  strictly speaking, the Ward-Takahashi and Slavnov-Taylor identities we discuss  actually mirror the way in which diffeomorphism invariance has been broken, and therefore often depend on the particular gauge choice. Many of the  identities  exist  because they involve the correlators of  gauge-variant fields, and hence cannot have a physically invariant meaning. Actual observables do not depend on any particular gauge choice, but in this paper we will not attempt to connect our gauge-variant correlators to any gauge-invariant observables like the statistical properties of the cosmic microwave background anisotropies. Although such a connection is relatively simple in linear perturbation theory, it is highly non-trivial beyond the linear order. 

At this point the Slavnov-Taylor identities could be viewed  as useful  checks on the validity of  intermediate cosmological perturbation theory calculations. In one-loop calculations of cosmological correlations, for instance, one needs to regularize the theory first. In some cases the regularization procedure may unintendedly break diffeomorphism invariance, and the resulting violation of the identities we derive can help  diagnose such violations. 

But, perhaps, the most important application of the Slavnov-Taylor identities is the derivation of ``consistency relations" between the different correlators of cosmological perturbations. To illustrate our methods, we derive consistency relations that follow  essentially from diffeomorphism invariance alone, although these lack the predictive power of other relations, and just seem to reflect the underlying redundancy associated with diffeomorphism invariance. The constraining power of diffeomorphism invariance  changes significantly when combined with an assumption about the analyticity of the correlators of the theory \cite{Berezhiani:2013ewa}. In this case, analyticity allows one to go beyond what appears to follow  merely  from gauge redundancies, allowing one to derive physically predictive  consistency relations in specific gauges, in the limit in which one of the field momenta approaches zero.  In that sense, the ensuing   relations are close relatives of  the constraints on the vertex function that guarantee the validity of the equivalence principle in  general relativity \cite{Brout:1966oea,DeWitt:1967uc}, which also follow from diffeomorphism invariance and analyticity around zero momentum transfer.

Some of the constraints that spatial diffeomorphism invariance imposes on the primordial perturbations have been recently discussed in \cite{Goldberger:2013rsa,Hinterbichler:2013dpa,Pimentel:2013gza}, and more specifically in references \cite{Berezhiani:2013ewa,Berezhiani:2014tda} by Berezhiani and Khoury. All these papers attempted to generalize or derive relations between correlation functions of cosmological perturbations that go back to the consistency condition originally  discussed by Maldacena in \cite{Maldacena:2002vr}.\footnote{Consistency relations following from diffeomorphisms have also been derived in the context of large scale structure---for more details, see for instance~\cite{Creminelli:2013mca,Horn:2014rta}.} Different arguments and symmetries have been used to derive such relations \cite{Creminelli:2004yq,Cheung:2007sv,Senatore:2012wy}, although it appears that an approximate conformal symmetry is the cleanest way to understand their origin \cite{Creminelli:2012ed,Hinterbichler:2012nm,Assassi:2012zq,Collins:2014fwa}. Our work is most similar to \cite{Berezhiani:2013ewa}, which we generalize to arbitrary gauges and also  extend to time diffeomorphisms.

The plan of the paper is as follows: In Section \ref{sec:Diffeomorphism Invariance} we explore the action of diffeomorphisms on cosmological perturbations, and establish how to calculate their expectation values. In Section  \ref{sec:Schwinger-Dyson Equations} we formulate identities for the connected correlators of the theory, while in Section \ref{sec:Slavnov-Taylor Identities} we derive equivalent identities on the one-particle-irreducible diagrams. We derive consistency relations between bispectra and power spectra in Section \ref{sec:Consistency Relations}, and finally conclude and summarize in Section \ref{sec:Summary and Conclusions}.

\section{Diffeomorphism Invariance}
\label{sec:Diffeomorphism Invariance}
Here, we are interested here in theories whose action is invariant under (infinitesimal) diffeomorphism transformations. The resulting equations of motion are then automatically covariant, so we take diffeomorphism invariance and general covariance to be synonymous.  We assume that these theories describe gravity coupled to a scalar field, so their action is of the general form
\begin{equation}\label{eq:S}
	S=S[g_{\mu\nu},\phi]\equiv \int_\mathcal{M} d^4 x \sqrt{-g}\,  \mathcal{L},
\end{equation}
where the Lagrangian density $\mathcal{L}$ depends on the metric $g_{\mu\nu}$, the scalar $\phi$, and their derivatives, and where the  integral runs over the spacetime manifold $\mathcal{M}$. Although we focus on a single scalar for simplicity, our results can easily be generalized to accommodate further scalar fields.

Under passive diffeomorphisms ${x^\mu\to x^\mu-\xi^\mu(x)}$  any tensor field $\bf{T}$ transforms as follows:
\begin{equation}\label{eq:delta diffs}
	\bf{T}\to \bf{T}+\pounds_\xi \bf{T},
\end{equation}
where $\pounds_\xi$ is the Lie-derivative along $\xi$. Hence,   if the Lagrangian density transforms like a scalar, $\mathcal{L}\to \mathcal{L}+\xi^\mu \partial_\mu \mathcal{L}$, the change in the action (\ref{eq:S}) under infinitesimal diffeomorphisms is given by
\begin{equation}\label{eq:S diff inv}
	\Delta S=\int_\mathcal{M} d^4 x \sqrt{-g} \, \nabla_\mu(\mathcal{L}\, \xi^\mu)=\int_{\partial \mathcal{M}} d^3 x \sqrt{\gamma} \, n_\mu \,  \mathcal{L} \, \xi^\mu,
\end{equation}
where $n^\mu$ is the normal to the boundary $\partial \mathcal{M}$ and $\gamma$ the determinant of its metric. Hence, in the diffeomorphism invariant theories we discuss here the action is actually only invariant up to boundary terms. 

\subsection{Cosmological Background}
Our goal is to constrain correlators of cosmological perturbations, that is, field fluctuations around a cosmological background. Therefore, we expand the metric and the scalar field as
\begin{equation}\label{eq:background + perturbations}
	\quad g_{\mu\nu}\equiv \bar{g}_{\mu\nu}+h_{\mu\nu}, \quad \phi \equiv \bar{\phi}+\varphi, 
\end{equation}
where $\bar{g}_{\mu\nu}$ and $\bar{\phi}$ are the background values of the metric and the scalar, and $h_{\mu\nu}$ and $\varphi$ are its fluctuations.  We choose our cosmological background to be that of a spatially flat  universe filled by a homogeneous scalar,
\begin{equation}\label{eq:FRW}
	d\bar{s}^2=a^2(\eta)\left[-d\eta^2+d\vec{x}\,^2\right],\quad
	\bar{\phi}=\bar\phi(\eta).
\end{equation}
Then, from equation (\ref{eq:delta diffs}),  the  perturbations around this background transform according to
\begin{equation}\label{eq:diffs}
	\Delta h_{\mu\nu}=g_{\mu\alpha}\partial_\nu \xi^\alpha+ g_{\alpha\nu} \partial_\mu \xi^\alpha+ \xi^\alpha \partial_\alpha g_{\mu\nu},
	\quad 
	\Delta \varphi=\xi^\alpha \partial_\alpha \phi,
\end{equation}
where $g_{\mu\nu}$ and $\phi$ are to be replaced by the corresponding expressions in equations (\ref{eq:background + perturbations}). Note that these transformations are valid to first order in $\xi$ (which we take to be infinitesimal), but to all orders in the fluctuations. In particular, diffeomorphisms act linearly (albeit non-homogeneously) on the field perturbations $h_{\mu\nu}$ and $\varphi$. 

\subsection{Cosmological Perturbations}
\label{sec:Cosmological Perturbations}

The isometry group of the background, those diffeomorphisms under which the background fields are invariant, plays a particularly important role in cosmological perturbation theory. Just as it is convenient to classify fields in Minkowski space according according to their transformation properties under the isometries of the Minkowski metric, it turns out to be convenient to classify cosmological perturbations in terms of their transformation properties under spatial translations and rotations.  We thus introduce a set of eleven  tensors $Q_{\mu\nu}{}^f(\vec{x};\vec{p})$ and  $Q^\varphi(\vec{x};\vec{p})$ that transform irreducibly under translations and rotations \cite{Bardeen:1980kt}. We list the components of these tensors in Appendix \ref{sec:Irreducible Tensors}.  What matters to us here is that we can expand the metric and scalar fluctuations in terms of these  tensors, 
\begin{equation}\label{eq:h real space}
	h_{\mu\nu}(\eta,\vec{x})=\sum_{f} \int d^3 p \, Q_{\mu\nu}{}^f (\vec{x};\vec{p})
	\,  f(\eta,\vec{p}), \quad
	\varphi(\eta,\vec{x})=\int d^3 p\,  Q^\varphi(\vec{x};\vec{p})\varphi(\eta,\vec{p}),
\end{equation}
where the sum over $f$ runs over the ten metric perturbation fields in momentum space
\begin{equation}
	f \in\{A,B, H_L, H_T,B_+, B_-, H_+,H_-, H_{++}, H_{--}\}.
\end{equation}
The fields $f = f, \varphi$ are eigenvectors of spatial translations by $\vec{a}$ [with eigenvalues $\exp(-i \vec{p}\cdot \vec{a})$], and spatial rotations by an angle $\theta$ around the $\vec{p}$ axis [with eigenvalues $\exp(-i m \theta)$ and $m=0$ for $f\in \{A,H_L,H_T,B,\varphi\}$ (scalars), $m=\pm1$ for $f\in\{B_\pm, H_\pm\}$ (vectors) and $m=\pm 2$ for $f\in\{H_{\pm\pm}\}$ (tensors)].
Conversely, given  arbitrary metric and scalar perturbations $h_{\mu\nu}(\eta,\vec{x})$ and $\varphi(\eta,\vec{x})$ we can determine the corresponding perturbation variables with the   projection operators  $Q_f{}^{\mu\nu}(\vec{p};\vec{x})$ and $Q_\varphi{}(\vec{p};\vec{x})$, whose components also we gather  in Appendix \ref{sec:Irreducible Tensors}. By definition, we thus have
 \begin{equation}
	f(\eta,\vec{p})=\int d^3 x \, Q_f{}^{\mu\nu}(\vec{p};\vec{x}) h_{\mu\nu}(\eta,\vec{x}),
	\quad
	\varphi(\eta,\vec{p})=\int d^3 x\,  Q_\varphi(\vec{p};\vec{x})\varphi(\eta,\vec{x}).
\end{equation}
Notice that the decomposition (\ref{eq:h real space}) of the metric fluctuations is equivalent to the following parametrization of the perturbed line element: 
\begin{eqnarray}
ds^2 &=& a^2 (\eta) \left\{-(1+2A)d \eta^2 + 2 \left( \frac{\partial_i B}{\sqrt{\nabla^2}} + B_i\right) dx^i d\eta \right. \\
&& \qquad \qquad \qquad \left. + \left[ \delta_{ij} (1 + 2 H_L) + 2 \left( \frac{\delta_{ij}}{3} - \frac{\partial_i \partial_j}{\nabla^2}  \right) H_T + 2 \frac{\partial_{(i} H_{j)}}{\sqrt{\nabla^2}} + H_{ij} \right] dx^i dx^j\right\}, \nonumber
\end{eqnarray}
where $B_i$ and $H_i$ are two transverse vectors with polarizations $B_\pm$ and $H_\pm$ respectively, while $H_{ij}$ is a traceless and transverse tensor with polarizations $H_{++}$ and $H_{--}$.

To simplify the notation it shall be convenient to simplify our equations by switching to deWitt notation \cite{DeWitt:1967ub}, in which Latin indices $a, b, \ldots$ collectively denote the type of field and its spacetime arguments, and the functional derivatives  are treated as partial derivatives ${\partial F/\partial f_a\equiv F{}^{,a}}$, and also denoted simply by $F^a$ where confusion is not likely.   Along the same lines, the index $\alpha$ shall denote both the components and the spacetime argument of the diffeomorphism parameter $\xi^\alpha(x)$. Indices in opposite locations thus imply both a sum over type of fields or parameter components, and an integral over spacetime arguments. 

For example, because diffeomorphisms are linear and inhomogeneous, we shall write equations (\ref{eq:diffs}) as 
\begin{equation} \label{eq:Delta}
	\Delta_a=(\mathcal{S}_a{}_\alpha
	+\mathcal{T}_a{}^b{}_\alpha\, f_b)\, \xi^\alpha,
\end{equation} 
where $\Delta_a$ is the change of the field $f_a$ under diffeomorphisms, and, in real space,  the non-vanishing components of the ``tensor" $\mathcal{S}_{a\alpha}$  are\begin{subequations}\label{eq:S real}
\begin{align}
	\mathcal{S}_{h_{\mu\nu}(x)}{}_{\xi^\alpha(y)}&=
	\left[\bar{g}_{\mu\alpha}\frac{\partial}{\partial x^\nu}+
	\bar{g}_{\alpha\nu}\frac{\partial}{\partial x^\mu}
	+\frac{\partial \bar{g}_{\mu\nu}}{\partial x^\alpha}\right]\delta^{(4)}(x-y),\\
	\mathcal{S}_{\varphi (x)}{}_{\xi^\alpha(y)}&=\frac{\partial\bar{\phi}}{\partial x^\alpha}\delta^{(4)}(x-y).
\end{align}
\end{subequations}
The free action for the perturbations is invariant under transformations with $\mathcal{T}_a{}^b{}_\alpha\equiv 0$, which is why we refer to (\ref{eq:S real}) as the transformation of the fields under ``linear diffeomorphisms." It readily follows from equation (\ref{eq:diffs}) that  for the  isometries of the background, namely, translations ($\bar{\xi}^\alpha=\delta^\alpha{}_i$) and rotations (${\bar{\xi}^\alpha=\epsilon^\alpha{}_{ij} x^j}$), the corresponding linear  transformations vanish, $\mathcal{S}_a{}_\alpha \, \bar{\xi}^\alpha=0$.  The non-vanishing components of diffeomorphism transformations linear in the field perturbations themselves, $\mathcal{T}_a{}^b{}_\alpha$, are  given by
\begin{subequations}\label{eq:T real}
\begin{align}
	\mathcal{T}_{h_{\mu\nu}(x)}{}^{h_{\rho\sigma}(y)}{}_{\xi^\alpha(z)}&=
	-\left[
	\delta_{\mu\alpha}{}^{\rho\sigma}
	\frac{\partial}{\partial z^\nu}
	+\delta_{\alpha\nu}{}^{\rho\sigma}
	 \frac{\partial}{\partial z^\mu}+
	\delta_{\mu\nu}{}^{\rho\sigma}
	 \frac{\partial}{\partial y^\alpha}\right]\delta^{(4)}(x-y)\delta^{(4)}(x-z),\\
	\mathcal{T}_{\varphi(x)}{}^{\varphi(y)}{}_{\xi^\alpha(z)}&=-
	\frac{\partial}{\partial y^\alpha}\delta^{(4)}(x-y)\delta^{(4)}(x-z).
\end{align}
\end{subequations}

In the above, a Kronecker delta function with 4 indices refers to a delta function symmetrized with respect to, say, both the upper indices.

Also, instead of the standard notation for the functional derivative $\delta \Delta f_a(x)/\delta \xi^\alpha(y)$, we shall write  the more compact expression
\begin{equation}\label{eq:linear transformation}
\Delta_{a\alpha}\equiv \frac{\partial \Delta_a}{\partial \xi^\alpha}\equiv \mathcal{S}_{a\alpha}
	+\mathcal{T}_a{}^b{}_\alpha f_b.
\end{equation}
In this notation, the transition between metric perturbation fields in real space, and the cosmological perturbations in Fourier space that we introduce in Appendix \ref{sec:Irreducible Tensors} amounts to a matrix multiplication. Denoting by $f_{\tilde{a}}$ the fields in Fourier space, and by $f_a$ those in real space, we have
\begin{equation}
	f_{\tilde{a}}=Q_{\tilde{a}}{}^a f_a, \quad
	f_a=Q_a{}^{\tilde{a}}f_{\tilde{a}},
\end{equation}
 with $Q_a{}^{\tilde{a}}\, Q_{\tilde{a}}{}^b=\delta_a{}^b$ and $Q_{\tilde{a}}{}^a\, Q_a{}^{\tilde{b}}=\delta_{\tilde{a}}{}^{\tilde{b}}$. Along the same lines, we can parameterize diffeomorphism transformations $\xi^\alpha$ in terms of its irreducible components $\xi^{\tilde{\alpha}}$, with
\begin{equation}
	\xi^{\tilde{\alpha}}=Q^{\tilde{\alpha}}{}_\alpha \xi^\alpha,
	\quad
	\xi^\alpha=Q^\alpha{}_{\tilde{\alpha}} \xi^{\tilde{\alpha}},
\end{equation} 
where the components of the transformations $Q$ are also listed in Appendix \ref{sec:Irreducible Tensors}. In this language, under diffeomorphisms the fields $f^{\tilde{i}}$ transform according to $f_{\tilde{a}}\to f_{\tilde{a}}+\Delta_{\tilde{a}}{}_{\tilde{\alpha}}\xi^{\tilde{\alpha}}$, where
 \begin{equation}\label{eq:Delta transform}
 	\Delta_{\tilde{a} \tilde{\alpha}}
	=Q_{\tilde{a}}{}^a \, \Delta_a{}_{,\alpha}\,  Q^\alpha{}_{\tilde{\alpha}}.
 \end{equation}
An advantage of this formalism is that it is covariant in field space. As long as the transformations between fields are linear, all our equations retain the same form, provided that the field tensors $\mathcal{S}$ and $\mathcal{T}$ are transformed appropriately. We list the components of $\mathcal{S}$ and $\mathcal{T}$ in the basis of the irreducible components in Appendix \ref{sec:Transformation under Diffeomorphisms}. 

\subsection{Expectation Values}
\label{sec:Expectation Values}
Primordial perturbations are characterized by the moments of the different metric perturbations at sufficiently early times. These moments are identified with equal time vacuum expectation values of the  corresponding product of  fields  in the quantum theory,
\begin{equation}
	\langle  \Pi_i f_{a_i}(\eta, \vec{x}_i)\rangle\equiv \langle 0_\mathrm{in}| \Pi_i f_{a_i}(\eta, \vec{x}_i)|0_\mathrm{in}\rangle.
\end{equation}
Therefore, to make predictions about the primordial perturbations we need to quantize the theory  and find a way to calculate expectation values of quantum fields. As far as the quantization is concerned, Fadeev and Popov   have argued that the canonical quantization of gravity  is equivalent to the covariant path-integral formulation, as long as one includes  appropriate gauge-fixing and ghost terms, and as long as one appropriately modifies the functional measure in the path integral \cite{Faddeev:1973zb}. The actual form of the path integral measure, however,  has been the subject of some controversy and does not appear to be settled  \cite{Unz:1985wq}. The author of the last reference, for instance, argues that the correct measure is 
\begin{equation}\label{eq:measure}
Dg\equiv \prod_{x\in \mathcal{M}} g^{00} \cdot g^{-1}\,  dg_{\mu\nu}(x), \quad
D\phi \equiv \prod_{x\in \mathcal{M}} (g^{00})^{1/2} \cdot g^{1/4}\, d\phi(x),
\end{equation}
and also suggests that, in spite of their appearance, both measures are invariant under diffeomorphisms.  We adopt the path integral formulation of quantum gravity here because it is better suited to handle local symmetries such as diffeomorphism invariance. The actual form of the measure is not important to us, as long as it is invariant under diffeomorphisms. This is  a requirement for the self-consistency of the theory, analogous to the demand that gauge theories be anomaly-free.

In order to calculate expectation values of fields at conformal time $\eta$ in the path integral approach, we  need to either double the number of fields \cite{Weinberg:2005vy}, or introduce a time contour $\mathcal{C}$ extending from the asymptotic past to  $\eta$  and back to the asymptotic past, ${\mathcal{C}\equiv (-\infty, \eta]\cup [\eta,-\infty)}$ \cite{CalzettaHu}. This last formulation is more convenient because it is formally analogous to  that of the in-out formalism, and because it allows us to work with a single set of fields.  In particular, the expectation value of a product of fields is simply 
\begin{equation}\label{eq:vev}
\langle 0_\mathrm{in}| \Pi_i f_{a_i}(\eta, \vec{x}_i)|0_\mathrm{in}\rangle=
\int Dh \, D\varphi \, D\omega \, \left[\Pi_i f_{a_i}(\eta, \vec{x}_i)\right] \exp\left(iS_\mathrm{tot}[h_\mathrm{\mu\nu},\varphi,\omega]\right),
\end{equation} 
where the functional integral runs over field configurations on the extended time contour $\mathcal{C}$, and we  have introduced the ghost fields $\omega$. The values of the fields at the endpoints of this contour, ${\partial \mathcal{C}= \{-\infty,-\infty\}}$ determine the state whose expectation value we are calculating. In the in-in formalism, the field configurations at both endpoints of the contour are identical, and hence, the boundary terms cancel and do not contribute to the change of the action under diffs. Therefore, any identity that follows from diffeomorphism invariance alone will apply to expectation values in arbitrary states. By shifting this contour by a small imaginary contribution, we can project onto the in-vacuum of the theory $|0_\mathrm{in}\rangle$.  

Naively, one may think that it is irrelevant whether we integrate over all metric and field configurations $g_{\mu\nu}$ and $\phi$, or just over its fluctuations $h_{\mu\nu}$ and $\varphi$, since they just differ by the given background values. But given the (somewhat uncertain) non-linear structure of the measure in equation (\ref{eq:measure}), such a shift may introduce fluctuation-dependent terms in the measure. Nevertheless, this has no impact on our analysis, as long as the measure for the new fields $h_{\mu\nu}$ and $\varphi$ remains diffeomorphism invariant, which, as we argued above, is a condition for the self-consistency of the theory. 

 The classical action for the perturbations is simply
\begin{equation}
	S_\mathrm{inv}[h_{\mu\nu},\varphi]\equiv S[\bar{g}_{\mu\nu}+h_{\mu\nu}, \bar{\phi}+\varphi],
\end{equation}
where the functional $S$ on the right-hand-side is diffeomorphism invariant, that is, satisfies equation (\ref{eq:S diff inv}).  Again,   $S_\mathrm{inv}$ is invariant  under the transformations of the perturbations (\ref{eq:diffs}) because in the in-in formalism there is no contribution from the boundary terms. 

The change of variables (\ref{eq:h real space}) casts the path integral in terms of the cosmological perturbations fields $f$. Because the transformation (\ref{eq:h real space}) is linear in the perturbations, the functional Jacobian is  field independent and has no impact on cosmological correlators. In particular, we can go back and forth between the representation of the fluctuations in terms of the fields $h_{\mu\nu}$ and $\varphi$ in real space, and the perturbations $f$ in Fourier space. On the other hand, a non-linear change of variables,   $h_{\mu\nu}(x)=F(\zeta_{\mu\nu}(x))$ would force us to introduce a field-dependent Jacobian in the path integral measure, which would amount to the additional term in the action
\begin{equation}
S_J=-i \Omega^{-1} \int d^4 x \log F'(\zeta_{\mu\nu}(x)),
\end{equation}
with a divergent constant $\Omega^{-1}=\delta^{(4)}(0)$. This would affect cosmological correlators, although only beyond tree level.  

\subsection{Gauge Fixing}
\label{sec:Gauge Fixing}

To render the functional integral (\ref{eq:vev}) well-defined, we have to introduce gauge-fixing and ghost terms into the action functional,
\begin{equation}\label{eq:S tot}
	S_\mathrm{tot}= S_\mathrm{inv}+S_\mathrm{gf}+S_\mathrm{gh}. 
\end{equation}
The actual form of the gauge-fixing terms is not particularly important, as long as they are \emph{not} invariant under the set of gauge symmetries under consideration.  In deWitt notation, if the gauge-fixing terms are taken to be of the form
\begin{equation}\label{eq:general fixing}
	\exp(iS_\mathrm{gf})=B[F_\beta(f_a)], 
\end{equation}
where $B$ is an arbitrary functional of its arguments, and the $F_\beta$ are a set of arbitrary local functions of the field perturbations $f_a$ (one function $F_\beta$ for each local symmetry), the only condition is that the matrix $F_\beta{}^{a} \Delta_{a\alpha}$ be invertible, which amounts to the functional $F_\beta$ not being invariant under any combination of infinitesimal diffeomorphisms.  We  focus here on tree-level calculations, so we shall  ignore the ghost fields, although they could be easily incorporated into our analysis.

\paragraph{Component Approach}

The conventional approach to gauge-fixing in cosmological perturbation theory is to impose conditions that  enforce the vanishing of a subset of the fields $f_g$, $g\in G$ 
\begin{equation}\label{eq:gf}
	\exp(iS_\mathrm{gf})= \prod_{g\in G} \delta(f_g).
\end{equation}
Because we are using deWitt notation, the index $g$ here runs over the fields that have been set to zero, and all their spacetime arguments.  Since diffeomorphisms are parameterized by four independent functions $\xi^\alpha$, we  need to specify four independent gauge fixing conditions, and we need to make sure that these conditions are not preserved by any infinitesimal diffeomorphism. Say, in longitudinal gauge we may   choose
\begin{equation}
	G_\mathrm{long}=\{B,H_T,H_+,H_-\}.
\end{equation}
Equations (\ref{eq:S fields}) then suffice to check that the condition $f_g =0\,  (g\in G_\mathrm{long})$ is not preserved by infinitesimal  diffeomorphisms.

The change in the total action (\ref{eq:S tot}) under diffeomorphisms plays a crucial role in the identities we derive below. By assumption, the classical action $S_\mathrm{inv}$ is invariant and thus does not contribute to the total change. 
There is also a simple way to calculate the change of the gauge fixing terms $S_\mathrm{gf}$, given a set of gauge fixing conditions ${f_g=0,\, g\in G}$. If in the absence of gauge-fixing conditions the action is gauge invariant, by definition it must be that
\begin{equation}
	0=\Delta S_{\mathrm{inv},\alpha}=
	 S_\mathrm{inv}{}^{a}\,\Delta_{a\alpha}.
\end{equation}
We now split the sum over the fields $f_g$ subject to the gauge condition $f_g=0$, and those fields $f_u$ which remain unconstrained, and impose the gauge-fixing conditions $f_g=0$  on the resulting equation,  
$S_\mathrm{inv}{}^{u}\,  \Delta_{u\alpha}|_{f_g=0}=
	 -S_\mathrm{inv}{}^{g}\,  \Delta_{g\alpha}|_{f_g=0}.
$
But this equation just states that the gauge-fixed action $S_\mathrm{inv}|_{g=0}=S_\mathrm{inv}+S_\mathrm{gf}$ changes by
\begin{equation}\label{eq:delta Sinv}
 (S_\mathrm{inv}+S_\mathrm{gf})_{,\alpha}=
 -S_\mathrm{inv}^{g}\, \Delta_{g\alpha}|_{g=0},	
\end{equation}
which completes our determination of the variation of the total action under diffeomorphisms. As an example consider the gauge fixing condition $\varphi=0$. Combination of equations (\ref{eq:diffs}) and (\ref{eq:delta Sinv}) implies that this condition breaks time, but preserves spatial diffeomorphisms.

\paragraph{Gauge-Fixing Terms}

The drawback of demanding that individual components of the field perturbations vanish is that the variation of the action under  broken diffeomorphisms in equation (\ref{eq:delta Sinv}) not only depends on the particular fields gauge-fixed to zero, but also on the actual invariant action $S_\mathrm{inv}$ of the theory.  In that case, it does not appear to  be possible to derive relations that only follow from diffeomorphism invariance, no matter what the specific action of the theory is. 

There is however a physically equivalent way to impose a gauge-fixing condition $f_g=0$ while preserving almost all of the symmetry of the action. Suppose  that we add to our action the gauge-fixing term
\begin{equation}\label{eq:M^2 gauge-fixing}
	S_\mathrm{gf}= - \frac{M^2}{2} \sum_{g\in G}    f_g^2,
\end{equation}
where $M$ is a constant that will be taken to infinity at the end of the calculation (the reader may think of this as a mass term for the field $f_g$.) This is to some extent analogous to the $R_\xi$ gauges employed in the quantization of non-abelian gauge symmetries. It amounts to a choice of a Gaussian $B$ in equation (\ref{eq:general fixing}) and a set of linear functions $F_\beta(f_a)\equiv \delta_{a\beta}$, where $\beta$ runs over the fields in $G$.  

For sufficiently large $M$,  the free propagator for the gauge-fixed fields $f_g$ and the remaining ``unconstrained" fields $f_u$ becomes 
\begin{equation}\label{eq:asymptotics}
	\langle f_{g_1} f_{g_2}\rangle_\mathcal{C}= - \frac{1}{M^2}\delta_{g_1 g_2}+\mathcal{O}(M^{-4}),  \quad 
	 \langle f_g f_u\rangle_\mathcal{C}=\mathcal{O}(M^{-2}).
\end{equation}
Hence,  in the limit $M\to \infty$, the fields $g$ decouple. The theory still has cubic and higher vertices containing the fields $f_g$, but their contributions to any diagram with no external heavy fields vanish because the internal line propagators approach zero as $M$ tends to infinity. Effectively, the theory is the same as if we had gauge-fixed $f_g\equiv 0$. If we keep $M$ finite, (\ref{eq:M^2 gauge-fixing}) remains a valid gauge-fixing term for appropriate choices of the fields $f_g$, but in this case, the massive fields $f_g$ are not decoupled from the theory.

\paragraph{Reduced Action}
\label{sec:Unitary Gauge}

Gauge symmetries enhance the invariance group of a theory, at the expense of introducing redundant degrees of freedom. In some cases, it is convenient to trade  back these gauge symmetries for a description of the theory that involves a smaller number of fields.  In many diffeomorphism invariant theories, such as general relativity coupled to a canonical scalar,  the metric components $h_{00}$ and $h_{0i}$ are auxiliary fields, and the matrix $S^{,ab}$ is non-singular for $a,b\in \{h_{00}, h_{0i}\}$. Therefore, if our gauge-fixed theory (\ref{eq:S tot})  belongs to the last class, we can  integrate these variables out perturbatively, 
\begin{equation}\label{eq:S R}
	\exp(i S_R[\varphi,h_{ij}])\equiv \int Dh_{00} \, Dh_{0i} \exp(i S_\mathrm{tot}).
\end{equation}
The resulting reduced action $S_R$ does not depend on the redundant variables $h_{00}$ and $h_{0i}$ any longer, and, as a result, it appears to have lost some of the original symmetries of $S_\mathrm{inv}$.  In fact, equation (\ref{eq:diffs}) for the spatial components of the metric
\begin{equation}
	\Delta h_{ij}=g_{i\alpha}\partial_j \xi^\alpha+g_{\alpha j} \partial_i \xi^\alpha+
\xi^\alpha \partial_\alpha g_{ij}, 
\end{equation}
implies that $S_R$ is invariant under spatial diffeomorphisms ($\alpha\neq 0$), but not under the original  time diffeomorphisms ($\alpha=0)$, basically because only under the former does $\Delta h_{ij}$  depend on the unconstrained variables $h_{ij}$ alone. The only exception consists of those  diffeomorphisms that amount to a spatially global time shift, $\xi^0=\xi^0(\eta)$, because in that case $\Delta h_{ij}$ does not involve the variables being integrated out.  But in any case, this apparent loss of invariance under diffeomorphisms is not fatal, among other things, because time diffeomorphisms have to be gauge-fixed (and hence broken) anyway. As it will become apparent below, at tree level it does not matter whether a symmetry has been  broken by gauge fixing terms or otherwise.

So far, we have simply integrated out the four auxiliary fields, but we have not fixed the gauge yet. In this context,  one usually fixes time-diffeomorphisms by imposing ``unitary gauge" $\varphi=0$. In unitary gauge the action is still invariant under spatial diffeomorphisms. We shall discuss different ways to gauge-fix the latter below.

\section{Schwinger-Dyson Equations for Connected Correlators}
\label{sec:Schwinger-Dyson Equations}

We are now ready to investigate the constraints diffeomorphism invariance places on the correlators of cosmological perturbations. For the sake of generality, we shall derive these identities  in an arbitrary gauge. Those terms in the  identities that involve variation with a field that has been gauged to zero should then be ignored.

Let us define the generator of connected correlators $W(J^a)$ by 
\begin{equation}\label{eq:W def}
	\exp (i W)=\int Df  \exp\left[i S_\mathrm{tot}+
	i J^a\,  f_a \right],
\end{equation}
where, again, fields are defined along the time contour $\mathcal{C}$ appropriate for the in-in formalism. Taking functional derivatives of $iW$ with respect to the currents $iJ^a$ and setting the latter to zero thus allows us to calculate contour-ordered correlators of arbitrary products of fields. Changing variables $f_a \rightarrow f_a+\Delta_{a}$ in equation (\ref{eq:W def}), and assuming that the measure is invariant under such an infinitesimal  transformation,   results  in  the master identity 
\begin{equation}\label{eq:W master}
J^a \left(\mathcal{S}_{a\alpha}+\mathcal{T}_a{}^b{}_\alpha\frac{\delta W}{\delta J^b}\right)=-W_{\Delta S,\alpha},
\end{equation}
where we have introduced  the generator of connected diagrams with an insertion of the change in the action under diffeomorphisms,
\begin{equation}
	W_{\Delta S,\alpha}=\frac{\int Df\,  
	(\delta \Delta S_\mathrm{tot}/\delta \xi^\alpha) \exp\left[i S_\mathrm{tot}+
	i J^a \, f_a\right]}{\int Df\,\exp\left[i S_\mathrm{tot}+
	i J^a \, f_a\right]},
\end{equation}
If the action $S_\mathrm{tot}$ is invariant under diffeomorphisms, $(\Delta S_\mathrm{tot})_{,\alpha}=0$, the generator $W_{\Delta S,\alpha}$  vanishes, but typically, the action contains gauge-fixing terms that break the symmetry, leading to a non-zero $W_{\Delta S,\alpha}$. In fact, the master identity above is valid no matter what the total action $S_\mathrm{tot}$ is. Any eventual change of the action under diffeomorphisms (or any transformation of the form (\ref{eq:linear transformation}) is then captured by  $W_{\Delta S,\alpha}$.

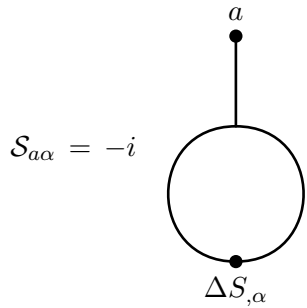
\begin{figure}
$\mathcal{S}_{a\alpha}\,=\,-i\,$
\subfigure
{
\parbox{20mm}{\begin{fmfgraph*}(20,30) 
\fmftop{i}
\fmfbottom{s2}
\fmfdot{i,s2}
\fmf{plain,tension=3}{i,s1}
\fmfpolyn{smooth,pull=?}{s}{2}
\fmflabel{$a$}{i}
\fmflabel{$\Delta S_{,\alpha}$}{s2}
\end{fmfgraph*}}
}

\caption{Diagrammatic representation of equation (\ref{eq:W,1}). Dots denote vertices, and circles the sum of all connected diagrams with the corresponding number of insertions. In particular, a  circle stands for $iW$, and a circle with a vertex insertion of $\Delta S_{,\alpha}$ stands for $W_{\Delta S_{,\alpha}}$. Each additional field $a$ connected to a circle then amounts to a functional derivative of the corresponding generator with respect to $i J^a$.}
\label{fig:W,1} 
\end{figure}

By taking functional derivatives of $W_{\Delta S,\alpha}$ with respect to the currents $i J^a$ we obtain the sum of all connected diagrams with the corresponding number of fields $f_a$ and a single insertion of $(\Delta S_\mathrm{tot})_{,\alpha}$. 
In  standard notation, letting $s(x)$ and $t^{\mu\nu}(x)$ denote the currents conjugate to $\varphi$ and $h_{\mu\nu}$, the master identity for the real space fields reads
\begin{equation}\label{eq:master W}
	s \frac{\partial}{\partial x^\alpha}\left(\bar{\phi}+\frac{\delta W}{\delta s(x)}\right)
	+t^{\mu\nu}\frac{\partial}{\partial x^\alpha}\left(\bar{g}_{\mu\nu}+\frac{\delta W}{\delta t^{\mu\nu}(x)}\right)
	-2\frac{\partial}{\partial x^\mu} \left[t^{\mu\nu} \left(\bar g_{\alpha\nu}+\frac{\delta W}{\delta t^{\alpha\nu}}\right)\right]
	 =-W_{\Delta S,\alpha}(x).
\end{equation}
Note that if we contract equation (\ref{eq:W master}) with the generator of a background isometry $\bar{\xi}^\alpha$, the inhomogeneous term $\mathcal{S}_{a\alpha} \, \bar{\xi}^\alpha$ drops out the equation. Typically, the gauge-fixing terms are chosen to respect the background isometries,   $(\Delta S_\mathrm{tot})_{,\alpha} \, \bar{\xi}^\alpha=0$, so  functional derivatives of the contracted equation express then the invariance of the correlation functions under such global transformations.

Equation (\ref{eq:master W})  (or (\ref{eq:W master})) captures the constraints imposed by diffeomorphism invariance on the connected diagrams of the theory and is one of the main results of this section. By taking functional derivatives of (\ref{eq:master W}) with respect to the currents $it^{\mu\nu}$ and $is$, we  derive relations between the correlators of cosmological perturbations.  If the gauge-fixing terms set some of the field perturbations $\varphi$ or $h_{\mu\nu}$ to zero, the generating functional does not depend on the associated conjugate currents, and the corresponding functional derivatives vanish. Similarly, if we are working with the reduced action (\ref{eq:S R}), the functional derivatives with respect to $t^{00}$ and $t^{0i}=t^{i0}$ can be set to zero, since these currents, and their conjugate fields, are not part of the theory. 

\begin{figure}
$i\mathcal{T}_a{}^c{}_\alpha$
\subfigure
{
\parbox{20mm}{\begin{fmfgraph*}(20,30) 
\fmftop{j}
\fmfbottom{k}
\fmfdot{k,j}
\fmf{plain,tension=3}{j,s1}
\fmfpolyn{smooth,pull=?}{s}{2}
\fmf{plain,tension=3}{s2,k}
\fmflabel{$b$}{j}
\fmflabel{$c$}{k}
\end{fmfgraph*}}
\!\!\!+\!\!\!
}
$i\mathcal{T}_b{}^c{}_\alpha$
\subfigure
{
\parbox{20mm}{\begin{fmfgraph*}(20,30) 
\fmftop{i}
\fmfbottom{k}
\fmfdot{i,k}
\fmf{plain,tension=3}{i,s1}
\fmfpolyn{smooth,pull=?}{s}{2}
\fmf{plain,tension=3}{s2,k}
\fmflabel{$c$}{k}
\fmflabel{$a$}{i}
\end{fmfgraph*}}
$={}$
}
\subfigure
{
\parbox{20mm}{\begin{fmfgraph*}(20,30)
\fmfstraight 
\fmftop{i,j}
\fmfbottom{k3}
\fmfdot{i,j,k3}
\fmfpolyn{smooth,pull=?}{k}{3}
\fmf{plain,tension=1.5}{i,k2}
\fmf{plain,tension=1.5}{j,k1}
\fmflabel{$a$}{i}
\fmflabel{$b$}{j}
\fmflabel{$\Delta S_{,\alpha}$}{k3}
\end{fmfgraph*}}
}

\caption{Diagrammatic illustration of equation (\ref{eq:W,2}). Same conventions as in Figure \ref{fig:W,1} apply.  Note that there is an implied sum and integral over the repeated dummy indices. The reader should be mindful of the different factors of $i$ that appear in the relation between connected diagrams and functional derivatives of $W$; for instance, the propagator is $-i W_{,ab}$.}
\label{fig:W,2}
\end{figure}
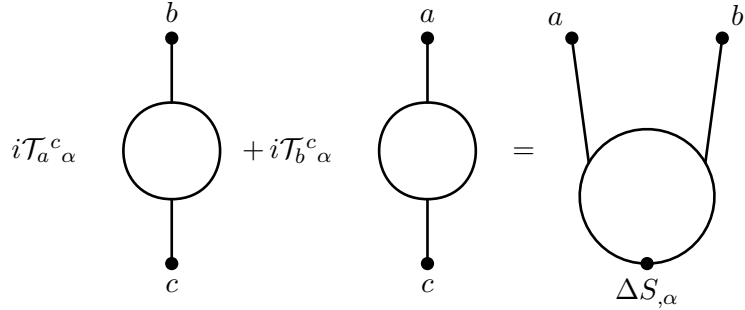

For instance, the simplest relation follows by just evaluating equation (\ref{eq:W master}) at zero currents,
\begin{equation}\label{eq:W,0}
	W_{\Delta S,\alpha}=0.
\end{equation}
This just states that the sum of all vacuum diagrams with an insertion of the vertex $\Delta S_{,\alpha}$ vanishes.  In some cases this would follow from translational invariance, although we have not made that assumption here. Taking one functional derivative of equation (\ref{eq:W master}) with respect to $J^a$ and setting the currents to zero then yields
\begin{equation}\label{eq:W,1}
	\mathcal{S}_{a\alpha}=-(W_{\Delta S,\alpha})_{,a},
\end{equation}
where we have assumed that the fields have zero expectation, $W_{,a}=0$ and we denote $\delta F/\delta J^a$ by $F_{,a}$ (again, where confusion is unlikely, we shall simply write $F_a$). The previous equation thus relates the sum of all connected diagrams with an insertion of $\Delta S_{,\alpha}$ and a single field $f_a$ to the inhomogeneous component of the change of $f_a$ under a diff transformation. We represent such a relation diagrammatically in Figure \ref{fig:W,1}.  Similarly, taking two functional derivatives of equation (\ref{eq:W master}) with respect to the currents yields the identity
\begin{equation}\label{eq:W,2}
\mathcal{T}_a{}^c{}_\alpha W_{cb}
+\mathcal{T}_b{}^c{}_\alpha W_{ca}=-(W_{\Delta S,\alpha})_{,ab},
\end{equation}
which relates the propagators of the theory to the sum of all connected diagrams with an insertion of $\Delta S_{,\alpha}$ and two fields, and is represented diagrammatically in Figure \ref{fig:W,2}. 

\begin{figure}
\subfigure
{
\parbox{20mm}{\begin{fmfgraph*}(20,30)
\fmfstraight 
\fmftop{i}
\fmfbottom{k2}
\fmfdot{i,k2}
\fmfpolyn{smooth,pull=?}{s}{2}
\fmf{plain,tension=2}{i,s1}
\fmfpolyn{smooth,pull=?}{k}{2}
\fmf{phantom ,tension=3}{s2,k1}
\fmflabel{$a$}{i}
\fmflabel{$\Delta S_{,\alpha}$}{k2}
\end{fmfgraph*}}
}+\,
\subfigure
{
\parbox{20mm}{\begin{fmfgraph*}(20,30)
\fmfstraight 
\fmftop{i}
\fmfbottom{s2}
\fmfpolyn{smooth,pull=?}{s}{2}
\fmfdot{i,s2}
\fmf{plain,tension=2}{i,s1}
\fmflabel{$a$}{i}
\fmflabel{$\Delta S_{,\alpha}$}{s2}
\end{fmfgraph*}}
}
\caption{The sum of all diagrams with insertions of $f_a$ and $\Delta S_{,\alpha}$, $\langle f_a \Delta S_{,\alpha}\rangle$, expressed in terms of sums of products of connected diagrams. We do not include vacuum-to-vacuum diagrams, which are  common factor to all of these, equal to one if the quantum state is normalized. Note that vacuum diagrams with a single field insertion vanish by assumption, $\langle f_a \rangle=0$, and that the vacuum diagrams with an insertion of $\Delta S_{,\alpha}$ vanish by equation (\ref{eq:W,0}). }
\label{fig:Z,1}
\end{figure}
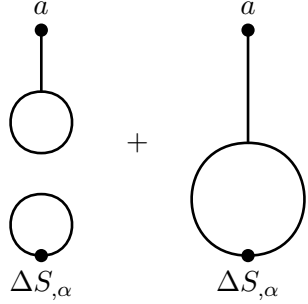

As we shall see, equations (\ref{eq:W,1}) and (\ref{eq:W,2}) are closely related to a family of relations known as Slavnov-Taylor  or Ward-Takahashi identities. To further illustrate their meaning, let us here elaborate on their connection with the Schwinger-Dyson equations. The latter reflect the fundamental theorem of calculus, namely, that the functional integral of a functional derivative vanishes. In particular, for any functional $F$ of the fields $f_a$ we have, in deWitt notation,
\begin{equation}\label{eq:DS start}
	\langle F^{c}\rangle+i \langle F S^{c}\rangle=0,
\end{equation}
where $S$ is the total action of the theory, and $\langle \cdots \rangle$ denotes the sum of \emph{all} diagrams (connected and disconnected) with the corresponding number of insertions. Equation (\ref{eq:DS start}) is also the statement that the  equations of motion $S^{c}=0$ of the classical theory hold in the quantum theory, modulo contact terms, for which the functional derivative $F^{c}$ is non-vanishing. Now, setting $F\equiv f_a \,\Delta_{c\alpha}$ in equation (\ref{eq:DS start}), and summing over $c$  results in the identity 
\begin{equation}\label{eq:SD}
	\langle \Delta_{a\alpha}+f_a \Delta_{c\alpha}{}^{,c}\rangle
	+i\langle f_a \Delta S_{,\alpha}\rangle=0.
\end{equation}
Because $\Delta_{a\alpha}$  is linear in the fields, taking Figure \ref{fig:Z,1} into account, and bearing in mind  that $\langle f_a\rangle=0$, this is nothing but equation (\ref{eq:W,1}). To arrive at this conclusion we also need to assume that $\Delta_{c\alpha}{}^{,c}=\mathcal{T}_c{}^c{}_\alpha=0$. This is again the statement that the integral of a derivative vanishes. As mentioned by deWitt in \cite{DeWitt:1967uc} it is also a condition for the internal consistency of the theory. Similarly, setting $F=f_a f_b \Delta_{c\alpha}$ in equation (\ref{eq:DS start}) and summing over $c$ we find,  
\begin{equation}
	\langle f_b \Delta_{a\alpha}\rangle
	+\langle f_a \Delta_{b\alpha}\rangle
	+\langle f_a f_b \Delta_{c\alpha}{}^{,c}\rangle
	+i\langle f_a f_b \Delta S_{,\alpha}\rangle=0.
\end{equation}
Taking Figure \ref{fig:Z,2} into account, and recalling equation (\ref{eq:W,0}), this becomes equation (\ref{eq:W,2}).  We can therefore think of equations (\ref{eq:W,1}) and (\ref{eq:W,2}) as consequences of the equations of motion in the quantum theory.

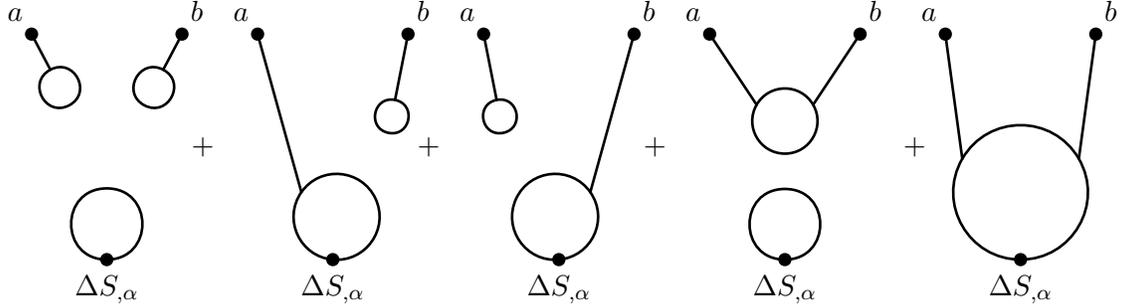
\begin{figure}[t]
\subfigure
{
\parbox{20mm}{\begin{fmfgraph*}(20,30) 
\fmftop{i,j}
\fmfbottom{k2}
\fmfdot{i,j,k2}
\fmfpolyn{smooth,pull=?}{s}{2}
\fmfpolyn{smooth,pull=?}{t}{2}
\fmf{phantom}{s1,x,t2}
\fmf{plain,tension=2}{i,s2}
\fmf{plain,tension=2}{j,t1}
\fmfpolyn{smooth,pull=?}{k}{2}
\fmf{phantom,tension=12}{x,k1}
\fmflabel{$\Delta S_{,\alpha}$}{k2}
\fmflabel{$a$}{i}
\fmflabel{$b$}{j}
\end{fmfgraph*}}
}+\,\,\,\,\, 
\subfigure
{
\parbox{20mm}{\begin{fmfgraph*}(20,30) 
\fmftop{i,j}
\fmfbottom{k3}
\fmfdot{i,j,k3}
\fmfpolyn{smooth,pull=?}{s}{2}
\fmf{plain,tension=1}{j,s1}
\fmf{phantom}{s2,k1}
\fmf{plain,tension=1}{i,k2}
\fmfpolyn{smooth,pull=?}{k}{3}
\fmflabel{$\Delta S_{,\alpha}$}{k3}
\fmflabel{$a$}{i}
\fmflabel{$b$}{j}
\end{fmfgraph*}}
}+\,\,\,\,\, 
\subfigure
{
\parbox{20mm}{\begin{fmfgraph*}(20,30) 
\fmftop{i,j}
\fmfbottom{k3}
\fmfdot{i,j,k3}
\fmfpolyn{smooth,pull=?}{s}{2}
\fmf{plain,tension=1}{i,s2}
\fmf{phantom}{s1,k2}
\fmf{plain,tension=1}{j,k1}
\fmfpolyn{smooth,pull=?}{k}{3}
\fmflabel{$\Delta S_{,\alpha}$}{k3}
\fmflabel{$a$}{i}
\fmflabel{$b$}{j}
\end{fmfgraph*}}
}+\,\,\,\,\, 
\subfigure
{
\parbox{20mm}{\begin{fmfgraph*}(20,30) 
\fmftop{i,j}
\fmfbottom{t2}
\fmfdot{i,j,t2}
\fmfpolyn{smooth,pull=?}{s}{3}
\fmf{plain}{i,s2}
\fmf{plain}{j,s1}
\fmflabel{$a$}{i}
\fmflabel{$b$}{j}
\fmf{phantom,tension=4}{s3,t1}
\fmfpolyn{smooth,pull=?}{t}{2}
\fmflabel{$\Delta S_{,\alpha}$}{t2}
\end{fmfgraph*}}
}\,\,\,\,\, +
\subfigure
{
\parbox{20mm}{\begin{fmfgraph*}(20,30)
\fmfstraight 
\fmftop{i,j}
\fmfbottom{k3}
\fmfdot{i,j,k3}
\fmfpolyn{smooth,pull=?}{k}{3}
\fmf{plain,tension=1.5}{i,k2}
\fmf{plain,tension=1.5}{j,k1}
\fmflabel{$a$}{i}
\fmflabel{$b$}{j}
\fmflabel{$\Delta S_{,\alpha}$}{k3}
\end{fmfgraph*}}
}
\caption{The sum of all diagrams with insertions of $f_a$, $f_b$ and $\Delta S_{,\alpha}$, $\langle f_a f_b \Delta S_{,\alpha}\rangle$, expressed in terms of sums of products of connected diagrams. Same comments as those in the caption to Figure \ref{fig:Z,1} apply. Hence, all but the last diagram on the right vanish.}
\label{fig:Z,2}
\end{figure}

\section{Slavnov-Taylor Identities for the Effective Action}
\label{sec:Slavnov-Taylor Identities}

In many cases, it is more convenient to restrict the properties of the one-particle irreducible diagrams of the theory, which are generated by the effective action.  The quantum effective action $\Gamma$ is the Legendre transformation of the generator of connected diagrams $W$,
\begin{equation}
\Gamma(\bar{f}_a)=W(J^a_*)-\bar{f}_a J^a_*,
\end{equation}
where the currents $J^a_*$ are defined by the condition
\begin{equation}
\frac{\delta W}{\delta J^a}\bigg|_{J = J_*}=\bar{f}_a,
\end{equation}
and $\bar{f}_a$ is the prescribed expectation value of the field $f_a$ (hence the bar, which we shall later drop for  simplicity.) The only difference here with respect to the in-out formalism is that, once more, time integrals run over the contour $\mathcal{C}$ we introduced in Section \ref{sec:Expectation Values}.

The generating functional $W$ does not depend on the currents conjugate to those fields that the gauge-fixing terms constrain to vanish, so the effective action does not depend on the corresponding field expectations. Therefore,  $\Gamma$ is  a functional of the prescribed expectations of the unconstrained perturbations alone.  Functional derivatives of $i\Gamma$ with respect to these fields  give the sum of all one-particle-irreducible diagrams with the corresponding number of external fields.  These one-particle-irreducible diagrams are then the building blocks from which one can calculate connected correlators,  by summing over tree diagrams whose vertices are determined by the corresponding functional derivatives of the effective action. 

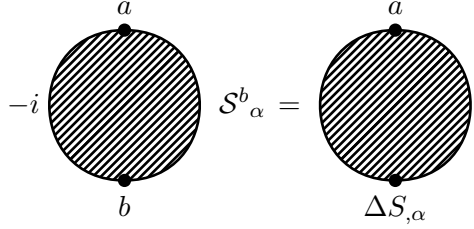
\begin{figure}
$-i$\subfigure
{
\parbox{20mm}{\begin{fmfgraph*}(20,20) 
\fmftop{j1}
\fmfbottom{j2}
\fmfdot{j1,j2}
\fmfpolyn{shaded,smooth,pull=?}{j}{2}
\fmflabel{$a$}{j1}
\fmflabel{$b$}{j2}
\end{fmfgraph*}}
}
$\mathcal{S}^b{}_\alpha\,=$
\subfigure
{
\parbox{20mm}{\begin{fmfgraph*}(20,20) 
\fmftop{j1}
\fmfbottom{j2}
\fmfdot{j1,j2}
\fmfpolyn{shaded,smooth,pull=?}{j}{2}
\fmflabel{$a$}{j1}
\fmflabel{$\Delta S_{,\alpha}$}{j2}
\end{fmfgraph*}}
}
\caption{Diagrammatic representation of equation (\ref{eq:Gamma,1}). Here, shaded circles denote sum of all one-particle-irreducible diagrams with the corresponding number of insertions. In particular, a shaded circle stands for $i\Gamma$, whereas a circle with a vertex labeled by $\Delta S_{,\alpha}$ stands for $\Gamma_{\Delta S,\alpha}$. Each additional field vertex $a$ denotes then a functional derivative of the corresponding quantum action with respect to $\bar{f}_a$.} 
\label{fig:Gamma,1}
\end{figure}

If the  action of a theory with fields $f_a$ changes by $\Delta S_{\mathrm{tot},\alpha}$ under an infinitesimal transformation $f_a\to f_a+\Delta_{a}$, where $\Delta_{a}$ is \emph{linear} in the fields like in equation (\ref{eq:Delta}), one can show (see e.g. \cite{ArmendarizPicon:2011ys}) that 
\begin{equation}\label{eq:Gamma master}
	\frac{\delta\Gamma}{\delta \bar f_a} (\mathcal{S}_{a\alpha}+\mathcal{T}_a{}^b{}_\alpha\bar{f}_b)=\Gamma_{{\Delta S},\alpha},
\end{equation}
where $\Gamma_{{\Delta S},\alpha}$  (note the missing factor of $i$) is the sum of all one-particle irreducible diagrams with an insertion of $\Delta S_{\mathrm{tot},\alpha}$. In particular, if the action is invariant, $\Delta S=0$, the previous equation states that linear symmetries are also symmetries of the effective action. If we contract equation (\ref{eq:Gamma master}) with an isometry unbroken by the gauge-fixing terms, the equation just expresses again the invariance of the effective action with respect those transformations, as before.   Equations relating the change of the effective action under a set of local transformations are generally known as Slavnov-Taylor identities, although they are often referred to as Ward-Takahashi identities too.  Adapting equation (\ref{eq:Gamma master}) to the standard notation, and dropping the bar from the arguments of the effective action  we obtain in real space
\begin{equation}\label{eq:PI Ward}
	\frac{\delta\Gamma}{\delta h_{\mu\nu}(x)}\frac{\partial g_{\mu\nu}}{\partial x^\alpha}
	+ \frac{\delta \Gamma}{\delta\varphi(x)}\frac{\partial \phi}{\partial x^\alpha}
	-2 \frac{\partial}{\partial x^\mu}\left(\frac{\delta\Gamma}{\delta h_{\mu\nu}(x)}g_{\alpha\nu}\right) = \Gamma_{\Delta S,\alpha}(x),
\end{equation}
where $g_{\mu\nu}$ and $\phi$ are the fields defined in equations (\ref{eq:background + perturbations}). This is our master identity for the effective action, which  holds for $\alpha=0$ (time diffeomorphisms) and $\alpha=i$ (spatial diffeomorphisms).  Again, if a certain set of fields are constrained to vanish by the gauge-fixing conditions, or they have been integrated out from the action,  the corresponding functional derivatives vanish. It is worth stressing that this master equation is valid at all orders in perturbation theory and for any gauge-fixing conditions. For spatial diffeomorphisms, and in the reduced formulation of the theory, essentially the same identity is derived in reference \cite{Berezhiani:2013ewa}. 

In some non-linear parameterizations of the metric perturbations, such as the one employed for instance in \cite{Maldacena:2002vr}, diffeomorphisms act non-linearly on the cosmological perturbations. In this case, equation (\ref{eq:Gamma master}) still holds at all orders in perturbation theory, provided that we truncate the action of diffeomorphisms on the fields of the theory to its linear components. In that case, $\Gamma_{\Delta S}$ on the right-hand-side will include the change in the action $\Delta S$ under these truncated linear diffeomorphisms, a change that would otherwise receive  contributions from the gauge fixing terms alone. At tree level this is a trivial consequence of the identity $\Gamma=S_\mathrm{tot}$ and  invariance of the classical action under diffeomorphisms,  $S_\mathrm{inv}^{,a} \, \Delta_{a\alpha}=0$. In particular, at tree level the Slavnov-Taylor equation
\begin{equation}
\frac{\delta \Gamma}{\delta \bar{f}_a}\,\Delta_{a\alpha}=\Gamma_{\Delta S,\alpha}
\end{equation}
holds even if $\Delta_{a\alpha}$ is a non-linear functional of the fields. 

\begin{figure}
\subfigure
{
\parbox{15mm}{\begin{fmfgraph*}(18,18) 
\fmftop{j1}
\fmfbottom{j3}
\fmfdot{j1,j2,j3}
\fmfpolyn{shaded,smooth,pull=?}{j}{3}
\fmflabel{$a$}{j1}
\fmflabel{$b$}{j2}
\fmflabel{$c$}{j3}
\end{fmfgraph*}}
}
$i\mathcal{S}_{c\alpha}\,+$
\subfigure
{
\parbox{20mm}{\begin{fmfgraph*}(20,20) 
\fmftop{j1}
\fmfbottom{j2}
\fmfdot{j1,j2}
\fmfpolyn{shaded,smooth,pull=?}{j}{2}
\fmflabel{$c$}{j2}
\fmflabel{$b$}{j1}
\end{fmfgraph*}}
}
$i\mathcal{T}_c{}^a{}_\alpha+$
\subfigure
{
\parbox{20mm}{\begin{fmfgraph*}(20,20) 
\fmftop{j1}
\fmfbottom{j2}
\fmfdot{j1,j2}
\fmfpolyn{shaded,smooth,pull=?}{j}{2}
\fmflabel{$c$}{j2}
\fmflabel{$a$}{j1}
\end{fmfgraph*}}
}
$i\mathcal{T}_c{}^b{}_\alpha=$
\subfigure
{
\parbox{35mm}{\begin{fmfgraph*}(18,18) 
\fmftop{j1}
\fmfbottom{j3}
\fmfdot{j1,j2,j3}
\fmfpolyn{shaded,smooth,pull=?}{j}{3}
\fmflabel{$b$}{j2}
\fmflabel{$a$}{j1}
\fmflabel{$\Delta S_{,\alpha}$}{j3}
\end{fmfgraph*}}
\!\!
}

\caption{Diagrammatic representation of equation (\ref{eq:Gamma,2}). Same conventions as in Figure \ref{fig:Gamma,1} apply.}
\label{fig:Gamma,2}
\end{figure}
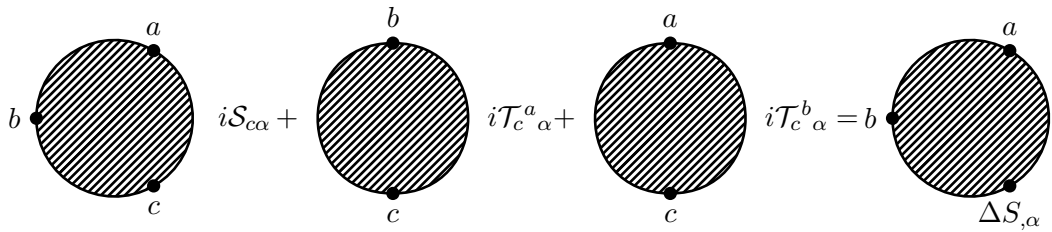

\subsection{Derivation of the Identities}

Taking functional derivatives of the effective action, and evaluating the latter at zero fields yields relations between the 1PI diagrams of the theory. For instance, taking one functional derivative of equation (\ref{eq:Gamma master}) and using that $\Gamma^a=0$ results in
\begin{equation}\label{eq:Gamma,1}
	\Gamma^{ba}\, \mathcal{S}_{b\alpha}=(\Gamma_{\Delta S,\alpha})^{,a},
\end{equation}
whereas taking two functional derivatives gives
\begin{equation}\label{eq:Gamma,2}
\Gamma^{cba}\mathcal{S}_{c\alpha}+\Gamma^{cb}\mathcal{T}_c{}^a{}_\alpha+\Gamma^{ca}\mathcal{T}_c{}^b{}_\alpha=(\Gamma_{\Delta S,\alpha})^{,ba}.
\end{equation}
These identities are represented diagrammatically in Figures \ref{fig:Gamma,1} and \ref{fig:Gamma,2}.

\begin{figure}
\subfigure
{
\parbox{20mm}{\begin{fmfgraph*}(20,30) 
\fmftop{i}
\fmfbottom{deltaS}
\fmfdot{i,deltaS}
\fmf{plain,tension=3}{i,j}
\fmf{plain,left}{j,deltaS}
\fmf{plain,right}{j,deltaS}
\fmflabel{$a$}{i}
\fmflabel{$\Delta S_{,\alpha}$}{deltaS}
\end{fmfgraph*}}
}
=
\subfigure
{
\parbox{20mm}{\begin{fmfgraph*}(20,30) 
\fmftop{i}
\fmfbottom{j2}
\fmfdot{i,j1,j2}	
\fmf{plain,tension=4}{i,m}
\fmf{plain,left,tension=2}{m,n}
\fmf{plain,right,tension=2}{m,n}
\fmf{plain,tension=4}{n,j1}
\fmfpolyn{shaded,smooth,pull=?}{j}{2}
\fmflabel{$a$}{i}
\fmflabel{$\Delta S_{,\alpha}$}{j2}
\end{fmfgraph*}}
}
\caption{The sum of all connected diagrams with an insertion of $\Delta S_{,\alpha}$ and a field $f_a$. The shaded blob indicates the sum of all 1PI diagrams with the given vertices, $\Gamma_{\Delta S,\alpha}$.}
\label{fig:W Delta S i}
\end{figure}
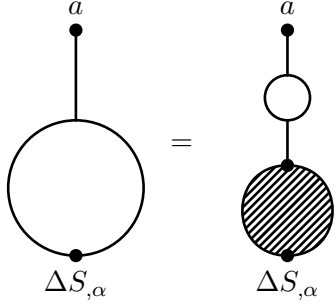

\begin{figure}
\subfigure
{
\parbox{30mm}{\begin{fmfgraph*}(25,30) 
\fmftop{i,j}
\fmfbottom{k3}
\fmfdot{i,j,k3}
\fmfpolyn{smooth,pull=?}{k}{3}
\fmf{plain,tension=2}{i,k2}
\fmf{plain,tension=2}{j,k1}
\fmflabel{$a$}{i}
\fmflabel{$b$}{j}
\fmflabel{$\Delta S_{,\alpha}$}{k3}
\end{fmfgraph*}}
}
=\, 
\subfigure
{
\parbox{30mm}{\begin{fmfgraph*}(25,30) 
\fmftop{i,j}
\fmfbottom{k3}
\fmfdot{i,j,k1,k2,k3}
\fmf{plain,tension=4}{i,m1}
\fmf{plain,left,tension=2}{m1,n1}
\fmf{plain,right,tension=2}{m1,n1}
\fmf{plain,tension=4}{n1,k2}
\fmf{plain,tension=4}{j,m2}
\fmf{plain,left,tension=2}{m2,n2}
\fmf{plain,right,tension=2}{m2,n2}
\fmf{plain,tension=4}{n2,k1}
\fmfpolyn{shaded,smooth,pull=?}{k}{3}
\fmflabel{$a$}{i}
\fmflabel{$b$}{j}
\fmflabel{$\Delta S_{,\alpha}$}{k3}
\end{fmfgraph*}}
}
+
\subfigure
{
\parbox{30mm}{\begin{fmfgraph*}(25,30) 
\fmftop{i,j}
\fmfbottom{s2}
\fmfdot{i,j,l1,l2,l3,s2}
\fmf{plain,tension=4}{i,m1}
\fmf{plain,left,tension=2}{m1,n1}
\fmf{plain,right,tension=2}{m1,n1}
\fmf{plain,tension=4}{n1,l2}
\fmf{plain,tension=4}{j,m2}
\fmf{plain,left,tension=2}{m2,n2}
\fmf{plain,right,tension=2}{m2,n2}
\fmf{plain,tension=4}{n2,l1}
\fmfpolyn{shaded,smooth,pull=?}{l}{3}
\fmf{plain,tension=4}{l3,s1}
\fmfpolyn{smooth,pull=?}{s}{2}
\fmflabel{$a$}{i}
\fmflabel{$b$}{j}
\fmflabel{$\Delta S_{,\alpha}$}{s2}
\end{fmfgraph*}}
}
\\
\caption{The sum of all connected diagram with an insertion of $\Delta S_\alpha$ and two fields $f_a$ and $f_b$. The shaded blobs indicates the sum of all 1PI diagrams with an insertion of $\Delta S_{,\alpha}$ and the number of fields indicated by the thick dots.}
\label{fig:W Delta S ij}
\end{figure}
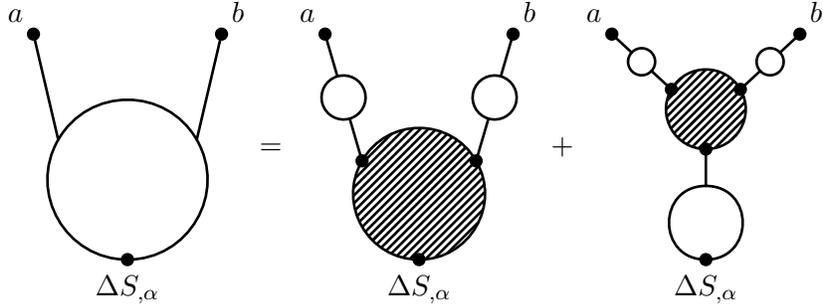

The reader may  wonder whether the identities that involve $\Gamma$ bear any relation to those obeyed by the generators of connected diagrams, $W.$  In fact, it is quite easy to see that both sets of identities are essentially the same. Compare for example equations (\ref{eq:W,1}) and (\ref{eq:Gamma,1}), and their corresponding diagrammatic representations in Figures \ref{fig:W,1} and \ref{fig:Gamma,1}. Because the sum of all connected diagrams with an insertion of $\Delta S_{,\alpha}$ and an external line is given by the diagrams in Figure \ref{fig:W Delta S i}, equation (\ref{eq:W,1}) just states that
\begin{equation}\label{eq:W constraint}
\mathcal{S}_{a\alpha}=-i\, (\Gamma_{\Delta S,\alpha})^{,b}\, (-i W_{ba}).
\end{equation}
Using the fact that the propagator $-iW_{ba}$ is just minus the inverse of the self-energy $i\Gamma^{ba}$ equation (\ref{eq:Gamma,1}) immediately follows. Similarly, equation (\ref{eq:W,2}) and Figure \ref{fig:W Delta S ij} imply that
\begin{equation}\label{eq:intermediate}
	\mathcal{T}_a{}^c{}_\alpha W_{cb}
+\mathcal{T}_b{}^c{}_\alpha W_{ca}=\Gamma_{\Delta S,\alpha}^ {cd} (-iW_{ca}) (-iW_{db})+i\Gamma^{ecd} (-iW_{ca})(-iW_{db}) \frac{1}{i}(W_{\Delta S,\alpha})_{,e}.
\end{equation}
Contracting left and right of this equation with two factors of the self-energy, and using equation (\ref{eq:W,1}) yields equation (\ref{eq:Gamma,2}).

Equation (\ref{eq:W,2}) relates cubic to quadratic terms in the effective action, and thus provides  constraints on the possible form of the cubic terms. These equations are analogous to the identities that relate the vertex for graviton emission by matter to the matter propagator, and ultimately enforce the equivalence principle in general relativity \cite{Brout:1966oea,DeWitt:1967uc}. If any of the fields appearing in these equations has been gauge-fixed to vanish, the corresponding term in the equation should be set to zero. The identities also hold in the reduced theory defined by equation (\ref{eq:S R}), provided  that functional derivatives of  $\Gamma$ with respect to $\bar{A}, \bar{B}$ and $\bar{B}_\pm$ are also set to zero. 

\subsection{Illustration}
\label{sec:Illustration}
As an application of these results, consider equation (\ref {eq:W constraint}) in a case in which the gauge fixing term is of the form (\ref{eq:M^2 gauge-fixing}). Then, at tree level, the effective action with an insertion of $\Delta S_{,\alpha}$ satisfies
\begin{equation}
(\Gamma_{\Delta S,\alpha})^{, a}= - M^2 \sum_g \delta_g{}^a\mathcal{S}_{g\alpha}.
\end{equation}
Therefore, inserting the last equation into (\ref {eq:W constraint}) we arrive at
\begin{equation}\label{eq:W constraint 2}
	\mathcal{S}_{a\alpha}= M^2\sum_g  \mathcal{S}_{g\alpha} W_{ga}.
\end{equation}
Suppose now the field $a$ is invariant under a particular linear diffeomorphism $\alpha$, $\mathcal{S}_{a\alpha}=0$, This is for instance the case for the gauge-invariant scalar perturbations introduced by Bardeen \cite{Bardeen:1980kt}. Then, if a particular  gauge-fixed field $g$ does transform under the same diffeomorphism, $\mathcal{S}_{g\alpha}\neq 0$, and also happens to be the only one appearing in the sum on the right hand side of (\ref{eq:W constraint 2}), it follows that $W_{ga}=0$. Therefore, at tree level, there is no correlation between gauge-invariant  and gauge-fixed perturbations. Note that we expect the fields $g$ to change under diffeomorphisms, because otherwise equation (\ref{eq:M^2 gauge-fixing}) would not be an appropriate gauge-fixing term.

\section{Consistency Relations from Diffeomorphisms}
\label{sec:Consistency Relations}

One of the main motivations for  the introduction of this formalism is the study of the extent to which diffeomorphism invariance constrains the properties of the  perturbations created during a scalar field driven inflationary stage. These primordial perturbations are conveniently characterized by the equal time expectation values of their products. For two fields we speak of spectra, and for three fields we speak of bispectra; these are the only quantities that are observationally relevant at this point. 

Building on the work of references  \cite{Goldberger:2013rsa,Hinterbichler:2013dpa}, the authors of \cite{Berezhiani:2013ewa}  considered the implications of spatial diffeomorphism invariance by formulating identities that relate the bispectrum of cosmological perturbations to their power spectrum, in the limit of a squeezed triangle. Similar relations had been derived earlier from the requirement of conformal invariance \cite{Creminelli:2012ed,Hinterbichler:2012nm,Assassi:2012zq}. Here, we extend the work of \cite{Berezhiani:2013ewa} to arbitrary diffeomorphisms and arbitrary gauges, not necessarily in the reduced formulation of the theory.  

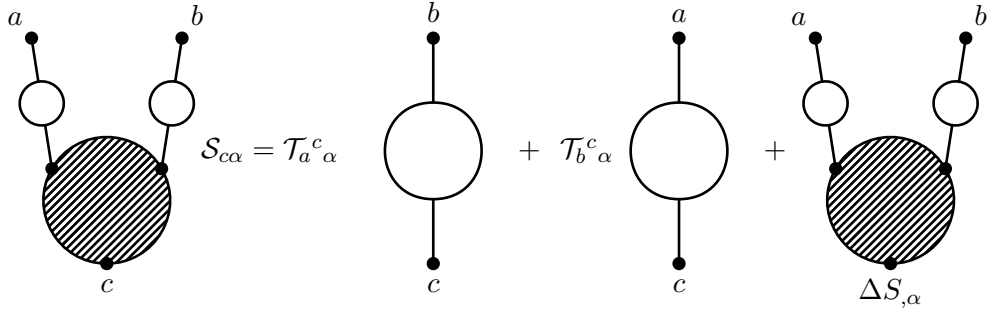
\begin{figure}
\subfigure
{
\parbox{20mm}{\begin{fmfgraph*}(20,30) 
\fmftop{i,j}
\fmfbottom{l3}
\fmfdot{i,j,l1,l2,l3}
\fmf{plain,tension=4}{i,m1}
\fmf{plain,left,tension=2}{m1,n1}
\fmf{plain,right,tension=2}{m1,n1}
\fmf{plain,tension=4}{n1,l2}
\fmf{plain,tension=4}{j,m2}
\fmf{plain,left,tension=2}{m2,n2}
\fmf{plain,right,tension=2}{m2,n2}
\fmf{plain,tension=4}{n2,l1}
\fmfpolyn{shaded,smooth,pull=?}{l}{3}
\fmflabel{$a$}{i}
\fmflabel{$b$}{j}
\fmflabel{$c$}{l3}
\end{fmfgraph*}}
}
$\mathcal{S}_{c\alpha}=\mathcal{T}_a{}^c{}_\alpha$
\subfigure
{
\parbox{20mm}{\begin{fmfgraph*}(20,30) 
\fmftop{j}
\fmfbottom{k}
\fmfdot{k,j}
\fmf{plain,tension=3}{j,s1}
\fmfpolyn{smooth,pull=?}{s}{2}
\fmf{plain,tension=3}{s2,k}
\fmflabel{$b$}{j}
\fmflabel{$c$}{k}
\end{fmfgraph*}}
+
}
$\mathcal{T}_b{}^c{}_\alpha$\!\!\!\!\!\!
\subfigure
{
\parbox{20mm}{\begin{fmfgraph*}(20,30) 
\fmftop{i}
\fmfbottom{k}
\fmfdot{i,k}
\fmf{plain,tension=3}{i,s1}
\fmfpolyn{smooth,pull=?}{s}{2}
\fmf{plain,tension=3}{s2,k}
\fmflabel{$c$}{k}
\fmflabel{$a$}{i}
\end{fmfgraph*}}
$+$
}
\subfigure
{
\parbox{20mm}{\begin{fmfgraph*}(20,30) 
\fmftop{i,j}
\fmfbottom{k3}
\fmfdot{i,j,k1,k2,k3}
\fmf{plain,tension=4}{i,m1}
\fmf{plain,left,tension=2}{m1,n1}
\fmf{plain,right,tension=2}{m1,n1}
\fmf{plain,tension=4}{n1,k2}
\fmf{plain,tension=4}{j,m2}
\fmf{plain,left,tension=2}{m2,n2}
\fmf{plain,right,tension=2}{m2,n2}
\fmf{plain,tension=4}{n2,k1}
\fmfpolyn{shaded,smooth,pull=?}{k}{3}
\fmflabel{$a$}{i}
\fmflabel{$b$}{j}
\fmflabel{$\Delta S_{,\alpha}$}{k3}
\end{fmfgraph*}}
}
\caption{Diagrammatic representation of equation (\ref{eq:consistency start}), the basis for the derivation of consistency relations in cosmological perturbation theory.}
\label{fig:consistency}
\end{figure}
  
Our goal is to  relate the bispectrum of cosmological perturbations to their power spectrum, with a gauge-fixing action of the form  (\ref{eq:M^2 gauge-fixing}).  Equation (\ref{eq:W,2}) appears to be the perfect starting point for such analysis, since its right hand side already contains almost what we are looking for: The sum of all connected diagrams with insertions of $f_a$, $f_b$ and $\Delta S_{,\alpha}$, the latter being known for a gauge-fixing term (\ref{eq:M^2 gauge-fixing}). Yet, since equation (\ref{eq:W,2}) contains contact terms that become singular in the limit of equal times, it is more convenient to start with the equivalent  identity (\ref{eq:intermediate}), which, when combined with equation (\ref{eq:W,1}) results in
\begin{equation}
	\Gamma^{cde} W_{da}W_{eb} \,\mathcal{S}_{c\alpha}=
	\mathcal{T}_a{}^c{}_\alpha W_{cb}
	+\mathcal{T}_b{}^c{}_\alpha W_{ca}+\Gamma_{\Delta S,\alpha}^ {de} W_{da} W_{eb}.
\end{equation} 
Diagrammatically, this equation can be represented as in Figure \ref{fig:consistency}. The left hand side of equation (\ref{eq:consistency start}) (or the equation in Figure \ref{fig:consistency}) is almost the sum of all connected diagrams with three external fields, since
$W_{abc}=\Gamma^{def}W_{da} W_{eb} W_{fc}$. To simplify the notation, we are going to think of $W_{ab}$ as a metric, which we can use to lower indices in field space. In this case, equation (\ref{eq:consistency start}) simplifies to
\begin{equation}\label{eq:consistency start}
\Gamma^c{}_{ab} \,\mathcal{S}_{c\alpha}=
	\mathcal{T}_a{}_b{}_\alpha
	+\mathcal{T}_b{}_a{}_\alpha +(\Gamma_{\Delta S,\alpha})_{ab}.
\end{equation}  
The reader should thus remember the natural position of the indices to determine whether an index has been lowered with the propagator. 

Equation (\ref{eq:consistency start}) is where we need to stop if we are not willing to make additional assumptions.  In order to proceed further, we shall work at tree level, where the analysis simplifies considerably, because the effective action $\Gamma$ is then just the same as the total (gauge-fixed) classical action $S_\mathrm{tot}$. If we further assume that the gauge-fixing term is of the form  (\ref{eq:M^2 gauge-fixing}) we know exactly what $\Delta S_{,\alpha}$ is, and because at tree level $(\Gamma_{\Delta S,\alpha})^{cd}=(\Delta S_{,\alpha})^{cd}$ it follows that 
\begin{equation}\label{eq:Gamma,alpha m n}
	(\Gamma_{\Delta S,\alpha})_{ab}= 
	- M^2 \sum_{g\in G}
	\left(\mathcal{T}_g{}^c{}_\alpha W_{ga} W_{cb}+\mathcal{T}_g{}^c{}_\alpha W_{ca} W_{gb}\right).
\end{equation}
Hence, if both external fields have a vanishing correlation with the massive fields $f_g$, the  last term on the right hand side of equation (\ref{eq:consistency start}) vanishes, regardless of the action of the theory and the particular diffeomorphism $\alpha$ involved. The latter is what happens for instance if the fields $f_g$ are scalars or vectors, and the fields $f_a, f_b$ are tensors. Note that this simplification occurs because of a global symmetry, namely,  invariance of the background under translations and rotations. In the meantime, we concentrate  on three-point functions that involve two tensor modes, for which the breaking term does not contribute under any circumstance. Later on we shall consider more general cases.

\subsection{Diffeomorphism Invariance}

With the term proportional to $\Gamma_{\Delta S_{,\alpha}}$ gone, we can focus on the irreducible vertex $\Gamma^c{}_{ab}$, which appears in the identity contracted with $\mathcal{S}_{c\alpha}$.   As seen from equation (\ref{eq:transverse S}), invariance under transverse diffeomorphisms ($\alpha=\pm$) constrains the vertices that include $B_{\pm}$ and $H_{\pm}$. Because equation  (\ref{eq:transverse S}) contains a time  derivative of a delta function, however, equation (\ref{eq:consistency start}) thus affects the time derivative of 1PI diagrams with an external vector $B_\pm$. It is hence not possible to translate such equation into an equation for connected correlators with a vector $B_\pm$, although such an equation would not be particularly relevant, since in the class of theories we are studying, vectors are redundant fields anyway. In the reduced formulation of the theory, for instance, the action $S_R$ defined in equation (\ref{eq:S R}) remains invariant under the  two transverse diffeomorphisms. Hence, in order to fix the gauge we can simply impose the condition $H_+=H_-=0$, thus eliminating  vectors from the theory altogether.  

Consider instead  longitudinal diffeomorphisms. In this case, the corresponding transformations in equation (\ref{eq:longitudinal S}) do not contain time derivatives of any field in the reduced formulation of the theory, in which $h_{00}$ and $h_{0i}$ have been integrated out, and the field $B$ is therefore absent (the term proportional to $\delta_f{}^B$ can be set to zero.) At this point, instead of working with the scalars $H_L$ and $H_T$, it is convenient to regard the effective action  in the scalar sector as a functional   of the two fields,
\begin{equation}\label{eq:L consistency gf}
	\Psi\equiv H_L+\frac{H_T}{3}, \quad H_L.
\end{equation}
Note that $\Psi$ is invariant under longitudinal diffeomorphisms in 
the linearized theory. Hence, by introducing these fields, we are dividing the two-dimensional scalar sector into a direction in field space that changes under linear longitudinal diffeomorphisms ($H_L$), and one which does not ($\Psi$). This division is to some extent arbitrary, since we may add any multiple of $\Psi$ to the gauge-variant direction, without changing the transformation properties under longitudinal diffs of the latter.  Clearly, in order to fix longitudinal diffeomorphisms we need to give a mass to $H_L$, which is the field that is not invariant under the symmetry,
\begin{equation}
	S_\mathrm{gf}= - M^2 \int d\eta\, d^3 p \, H_L(\eta,\vec{p}) H_L(\eta,-\vec{p}).
\end{equation}
 In terms of $\Psi$ and $H_L$, and in unitary gauge, equation (\ref{eq:transverse S}) simply reduces to  
\begin{equation}\label{eq:longitudinal S prime}
	\mathcal{S}_{f(\eta_1,\vec{p}_1) \,\xi^L(\eta_2,\vec{p}_2)}=
	\frac{p_1}{3} \delta(\eta_1-\eta_2) \delta(\vec{p}_1-\vec{p}_2)  \delta_f{}^{H_L}.
\end{equation}
Equation (\ref{eq:consistency start}) thus constrains 1PI diagrams that contain the scalar field $H_L$, the gauge-variant direction in the scalar sector.

To obtain an equation that involves the connected correlators, we  need to contract equation (\ref{eq:consistency start}) with a scalar propagator. According to the discussion of Section \ref{sec:Illustration}, invariance under longitudinal diffeomorphisms ($\alpha=L$) implies that there is no correlation between the gauge-invariant field $\Psi$ and the gauge-fixed field $H_L$, $W_{H_L\Psi}=0.$ Hence, the propagator  in the scalar sector is diagonal in the fields $H_L$ and $\Psi$.  Because the propagator is diagonal, and equation (\ref{eq:consistency start}) involves a vertex with a field $H_L$, we  just multiply the left and right hand side of equation (\ref{eq:consistency start}) by $W_{\tilde{H}_L H_L}$. Recall that if  $f_a$ and $f_b$ are tensors, the term that involves $\Gamma_{\Delta S}$ does not contribute. Hence,  setting $a$ and $b$ to be tensors with respective helicities $\sigma_2$ and $\sigma_3$, and integrating over the undisplayed time and momentum variables we finally obtain
\begin{multline}\label{eq:L consistency}
\frac{p_1}{3}\frac{W_{H_L H_{\sigma_2} H_{\sigma_3}}(\eta,\vec{p}_1;\eta,\vec{p}_2;\eta,\vec{p}_3)}{\bar{W}_{H_L H_L}(\eta,\vec{p}_1)}= \bigg\{ \bigg[
	2 p_j^1 Q_{H_{\sigma_3}}{}^{ij}(\vec{p}_3)Q_{ik}{}^{H_{\sigma_2}}(-\vec{p}_2) \hat{p}_1^k\\
	+Q_{H_{\sigma_3}}{}^{ij} (\vec{p}_3)
	Q_{ij}{}^{H_{\sigma_2}}(-\vec{p}_2) \, \vec{p}_2\cdot \hat{p}_1\bigg]
	\bar{W}_{H_{\sigma_2}H_{\sigma_2}}(\eta,\vec{p}_2)
	+2\leftrightarrow 3\bigg\}\frac{\delta(\vec{p}_1+\vec{p}_2+\vec{p}_3)}{(2\pi)^{3/2}},
\end{multline}
where we define the power-spectrum of an arbitrary variable $f$ by
\begin{equation}
	\langle f(\eta,\vec{p}) f(\eta,\vec{p}\,')\rangle \equiv		-i\bar{W}_{ff}(\eta,\vec{p}) \delta(\vec{p}+\vec{p}\,'),
\end{equation}
and the components of the projection tensors for tensor perturbations are listed in Appendix \ref{sec:Irreducible Tensors}. 
Equation (\ref{eq:L consistency}) therefore relates the three-point function of cosmological perturbations to their power spectra. It is the consistency relation that follows from the original invariance under longitudinal diffeomorphisms. Note that it is valid for all scalar momenta, and not only in the soft limit $\vec{p}_1\to 0$.  As should be manifest from our derivation, it applies only at tree level  and in the reduced formulation of the theory, with a  gauge fixing term (\ref{eq:L consistency gf}) that gives $H_L$ an arbitrary (but finite) mass. Other than that it only relies on the invariance of the theory under spatial diffeomorphisms and the isometries of a cosmological background.  The consistency relation does not explicitly contain $M^2$, although the power spectra and the three-point function implicitly depend on that quantity.  It is also important to realize that the gauge in which this consistency relation holds is not one of the conventional gauge choices used in cosmological perturbation theory. By giving a finite mass term to a scalar variable, we are not eliminating any scalar from the theory. Hence,  the scalar sector here consists of two dynamical fields ($\Psi$ and $H_L$), rather than one, as in the standard $\zeta$-gauge, in which $H_T\equiv 0$. If we had simply set $H_T$ to zero, we would have lost  the ability to calculate $\Gamma_{\Delta S}$.

Diffeomorphism invariance  also constrains the expectation of a product of three  scalar perturbations, or the product of two scalars and a tensor.  In this case, however, the last term on the right hand side of equation (\ref{eq:consistency start})  contributes a non-vanishing correction proportional to $M^2$. As a result, the ensuing consistency relation becomes explicitly gauge-dependent. Hence, we shall not write down the corresponding consistency relation here, although it can be easily derived from the previous equations. A consistency relation for time diffeomorphisms can be derived along the same lines. 

\subsection{Analyticity}

As we have seen, diffeomorphism invariance alone  constrains the cubic vertices of the theory only along gauge-variant directions in field space.  As shown in reference \cite{Berezhiani:2013ewa}, however, additional analyticity properties allow us to extend these constraints to the full cubic vertex itself, in the limit in which one of the momenta approaches zero.  

\subsubsection{Spatial Diffeomorphisms}
\label{sec:Spatial Diffeomorphisms}

\paragraph{Unitary Gauge}
To see how this works, it is going to be useful to consider   
the sum of all diagrams with insertions of  $h_{ij}(\eta_1,\vec{p}_1)$ and two arbitrary fields $f_2(\eta,\vec{p}_2)$ and $f_3(\eta,\vec{p}_3)$, with the propagator of $h_{ij}$ stripped off, and the overall momentum-conserving delta function omitted,
\begin{equation}\label{eq:mixed vertex}
	\bar\Gamma^{ij}{}_{f_2f_3}(\eta_1,\vec{p}_1;\eta,\vec{p}_2;\eta)\times\delta^{(3)}(\vec{p}_2+\vec{p}_3-\vec{p}_1)\equiv \Gamma^{f_1}{}_{f_2f_3} Q_{f_1}{}^{ij}(\eta_1,\vec{p}_1).
\end{equation}
We consider an insertion of the metric perturbation $h_{ij}$, rather than a helicity eigenvector $f_1$, because the decomposition into irreducible representations obscures the analyticity properties of the vertex. We also define the propagator of the fields with a momentum-conserving delta function stripped off, which in the case of coincident times defines the power spectrum, 
\begin{equation}
	W_{f_1f_2}(\eta_1,\vec{p}_1;\eta_2,\vec{p}_2)
	\equiv
	\bar{W}_{f_1f_2}(\eta_1,\vec{p}_1;\eta_2)\times \delta(\vec{p}_1+\vec{p}_2).
\end{equation}

We work in the reduced formulation of the theory, in a gauge  in which $\varphi\equiv 0$ and the breaking term is 
\begin{equation}\label{eq: M H_T}
S_\mathrm{gf}= - \frac{M^2}{2}\int d\eta\,  d^3p \, \left[H_T(\eta,\vec{p})H_T(\eta,-\vec{p})+H_+(\eta,\vec{p})H_+(\eta,-\vec{p})+H_-(\eta,\vec{p})H_-(\eta,-\vec{p})\right].
\end{equation} 
We take the limit $M\to \infty$ at the end of the calculation, which decouples both the scalar $H_T$ and the two vectors $H_{\pm}$. This implies that we assume the fields $f_2$ and $f_3$ to stand for the remaining light fields $H_L$ or $H_{\pm\pm}$, but not for any of the massive fields.  Then, from equation (\ref{eq:consistency start}), invariance under spatial diffeomorphisms $\xi^j$ implies that $\bar\Gamma^{ij}{}_{f_2f_3}$ obeys the equation
\begin{multline}\label{eq:ST mixed vertex}
2 a_1^2 p_i^1 \bar\Gamma^{ik}{}_{f_2f_3}\delta_{kj}
-
\Bigg[\Bigg(2p^1_k  Q_{f_2}{}^{ik}(\vec{p}_2) Q_{ij}{}^{f_3}(\vec{p}_2-\vec{p}_1) +(p^2_j-p^1_j)Q_{f_2}{}^{ik}(\vec{p}_2)Q_{ik}{}^{f_3}(\vec{p}_2-\vec{p}_1) \Bigg)
\bar W_{f_3 f_3}(\eta,\vec{p}_2-\vec{p}_1)\\
+\Bigg(2p^1_k  Q_{f_3}{}^{ik}(\vec{p}_1-\vec{p}_2) Q_{ij}{}^{f_2}(-\vec{p}_2) -p^2_j Q_{f_3}{}^{ik}(\vec{p}_1-\vec{p}_2)Q_{ik}{}^{f_2}(-\vec{p}_2) \Bigg)
\bar W_{f_2 f_2}(\eta,-\vec{p}_2)
\Bigg]\frac{\delta(\eta-\eta_1)}{(2\pi)^{3/2}}
=-i(\bar\Gamma_{\Delta S,j})_{f_2f_3},
\end{multline}
where we have defined
\begin{equation}
	(\bar\Gamma_{\Delta S,j})_{f_2f_3}(\eta_1,\vec{p}_1;\eta,\vec{p}_2;\eta)\delta^{(3)}(\vec{p}_2+\vec{p}_3-\vec{p}_1)\equiv\left(\frac{\delta\Gamma_{\Delta S}}{\delta \xi^j(\eta_1,\vec{p}_1)}\right)^{\tilde{f}_2\tilde{f}_3}
W_{\tilde{f}_2f_2}
	W_{\tilde{f}_3f_3}.
\end{equation}
Note that because in the limit $M\to \infty$ the propagator is diagonal in field space, there is no need to sum over the repeated indices $f_2$ and $f_3$ in equation (\ref{eq:ST mixed vertex}). On the other hand, because the breaking term is proportional to $M^2$, one needs to be careful with terms in the propagator that only decay like $1/M^2$ when one deals with $(\bar\Gamma_{\Delta S,j})_{f_2 f_3}$. 

We can  constrain the  components of $\bar\Gamma^{ij}{}_{f_2f_3}$  if we assume it to be analytic for momenta $\vec{p}_1$ in the vicinity of zero. As argued in \cite{Berezhiani:2013ewa}, this is a non-trivial assumption even at tree-level because gravitons are massless particles and we are working in the reduced formulation of the theory, in which $h_{00}$ and $h_{0i}$ have been integrated out. Nevertheless, if the assumption holds, we can solve for  $\bar{\Gamma}^{ij}{}_{f_2f_3}$ as  power series in the components of $\vec{p}_1.$  Say, at zeroth order in $\vec{p}_1$, we find that the equation is satisfied provided that 
\begin{equation}\label{eq:consistency zeroth}
	(\bar{\Gamma}_{\Delta S,j})_{f_2f_3}(\eta_1,\vec{p}_1=0;\eta,\vec{p}_2;\eta)=0.
\end{equation}
We shall verify this property below. At first order we obtain then the unique solution
\begin{multline}\label{eq:K0 mn pq}
2a_1^2  \bar{\Gamma}^{ik}_{(0)}{}_{f_2f_3} \delta_{kj}=
	-i\frac{\partial(\bar{\Gamma}_{\Delta S,j})_{f_2f_3}}{\partial p^1_i}+\frac{\delta(\eta-\eta_1)}{(2\pi)^{3/2}}
	\Bigg[-\delta^i{}_j \bar{W}_{f_2f_3}(\eta,\vec{p}_2)-p^2_j \frac{\partial \bar W_{f_2f_3}(\eta,\vec{p}_2)}{\partial p^2_i}\\
	+2Q_{f_2}{}^{ik}(\vec{p}_2)Q_{kj}{}^{f_3}(\vec{p}_2)\bar{W}_{f_3 f_3}(\eta,\vec{p}_2)
	+2Q_{f_3}{}^{ik}(-\vec{p}_2)Q_{kj}{}^{f_2}(-\vec{p}_2)\bar{W}_{f_2 f_2}(\eta,-\vec{p}_2)
	\Bigg],
\end{multline}
Although this is not immediately apparent, it is straight-forward to check that for  a  rotationally-invariant state of the perturbations the right-hand side is always symmetric in~$ij$. For instance, $p_j^2\, \partial \bar{W}_{f_2f_3}/\partial p^2_i$ is symmetric if $\bar{W}_{f_2f_3}$  only depends on the magnitude of the vector $\vec{p}_2$.   

Along the same lines, one can derive the solution of equation (\ref{eq:ST mixed vertex}) at higher orders in the momentum $\vec{p}_1$. At first order the solution is again unique, but the proliferation of indices makes its manipulation rather cumbersome beyond the hard scalar case, in which $f_2=H_L$, $f_3=H_L$.   At yet higher orders the solution is not unique because equation (\ref{eq:ST mixed vertex}) only constrains the longitudinal component of the vertex.

In order to proceed, we need to determine $(\bar\Gamma_{\Delta S,j})_{f_2f_3}$.  With a breaking term of the form (\ref{eq: M H_T}) the variation of the quadratic part of the action under  spatial diffeomorphisms ${\alpha=\xi^j(\eta_1,\vec{p}_1)}$ becomes
\begin{multline}
	i(\Gamma_{\Delta S,j})_{f_2f_3}=
	\frac{M^2}{(2\pi)^{3/2}}
	\Bigg\{
	\bar{W}_{f_2 H_T}(\eta,\vec{p}_2;\eta_1)\bar{W}_{f_3 \tilde{f}_3}(\eta,\vec{p}_3;\eta_1)\Big[2 p_k^1 Q_{H_T}{}^{ik}(\vec{p}_2) Q_{ij}{}^{\tilde{f}_3}(-\vec{p}_3)\\
	-p^3_j Q_{H_T}{}^{ik}(\vec{p}_2)Q_{ik}{}^{\tilde{f}_3}(-\vec{p}_3)\Big]+2\leftrightarrow 3\Bigg\}\delta^{(3)}(\vec{p}_2+\vec{p}_3-\vec{p}_1).
\end{multline}

It is easy to check that $(\bar\Gamma_{\Delta S,\xi^k})_{f_2f_3}$ vanishes at $\vec{p}_1=0$, as required by condition (\ref{eq:consistency zeroth}). 
In fact, as we also stated above,  for two hard tensors $f_2$ and $f_3$ the breaking term vanishes at all momenta and can be therefore ignored.  On the other hand, this simplification does not generically occur when the two fields $f_2$ and $f_3$ are scalars. If, for instance, $f_2=H_L^{(2)}(\eta,\vec{p}_2)$ and $f_3=H_L^{(3)}(\eta,\vec{p}_3$),  the first derivative of $(\bar\Gamma_{\Delta S,j})_{f_2f_3}$ equals
\begin{multline}\label{eq:delta Gamma L}
	\frac{\partial
	(\bar\Gamma_{\Delta S,j})_{H_L^{(2)}H_L^{(3)}}}{\partial p^1_i}\propto - M^2 
	Q_{H_T}{}^{ik}(\vec{p}_2) Q_{kj}{}^{H_L}
	\big[
	\bar{W}_{H_L H_T}(\eta,\vec{p}_2; \eta_1)
	\bar{W}_{H_L H_L}(\eta,-\vec{p}_2;\eta_1)\\
	+ \bar{W}_{H_L H_T}(\eta,-\vec{p}_2; \eta_1)
	\bar{W}_{H_L H_L}(\eta,\vec{p}_2;\eta_1)\big] ,
\end{multline}
where we have used that $Q_{H_T}{}^{ij}(-\vec{p}_2) Q_{ij}{}^{H_L}(\vec{p}_3)\equiv 0$. Note that $\bar{W}_{H_L H_T}$ is proportional to $1/M^2$, so the right hand side remains finite  in the limit  $M\to \infty$. Since the finite limit of $M^2 \bar{W}_{H_L H_T}$ depends on details of the theory, we cannot hence determine the contribution of this term from symmetry arguments alone. There is however an exception. If we were interested in correlation functions with a soft scalar, we would need to contract the vertex with $Q_{ij}{}^{H_L}$, since $\bar\Gamma^{H_L}{}_{H_L H_L}\equiv \bar{\Gamma}^{ij}{}_{H_L H_L}Q_{ij}{}^{H_L}$.  Because the right-hand side of equation (\ref{eq:delta Gamma L}) is traceless, the contribution of the breaking term would then vanish. Similarly, one can also check that the breaking term does not contribute to $\Gamma^{H_L}{}_{H_L H_{\pm\pm}}$.  We summarize the combination of fields for which the symmetry-breaking term can be discarded at zeroth order in $\vec{p}_1$ in table \ref{table:combinations unitary}.

\begin{table}  
\begin{center}
\begin{tabular}{c | c | c}
$f_1$ & $f_2$ & $f_3$\\ \hline
$H_L$ & $H_L$ & $H_L$ \\
$H_L$ & $H_L$ & $H_{\pm\pm}$ \\ 
$H_L$ & $H_{\pm\pm}$ & $H_{\pm\pm}$\\
$H_{\pm\pm}$ & $H_{\pm\pm}$ & $H_{\pm\pm}$
\end{tabular}
\end{center}
\caption{The different combination of fields for which the consistency relation (\ref{eq:spatial diffs consistency}) in unitary gauge holds. The field $f_1$ carries the soft momentum, whereas the two hard fields $f_2$ and $f_3$ can be interchanged.
\label{table:combinations unitary}} 
\end{table}

The  offshoot of the previous analysis is that at zeroth order in $\vec{p}_1$, and for the combination of fields listed in table \ref{table:combinations unitary}, the mixed vertex $\bar\Gamma^{f_1}{}_{f_2f_3}$ is determined by  symmetry alone. We are then just a step away from the consistency relation for spatial diffeomorphisms.   Convolving  $\bar\Gamma^{f_1}{}_{f_2f_3}$ with the propagator of $f_1$ we finally arrive at
\begin{subequations}\label{eq:spatial diffs relations}
\begin{multline}\label{eq:spatial diffs consistency}
	\frac{\bar{W}_{f_1 f_2f_3}(\eta,\vec{p}_1;\eta,\vec{p}_2;\eta)}{\bar{W}_{f_1f_1}(\eta,\vec{p}_1;\eta)}=
	\frac{Q_{ik}{}^{f_1}(-\vec{p}_1)\delta^{kj}}{2(2\pi)^{3/2}}\Bigg[-\delta^i{}_j\bar{W}_{f_2f_3}(\vec{p}_2)-p^2_j \frac{\partial\bar{W}_{f_2f_3}}{\partial p^2_i}+2 Q_{f_2}{}^{ik}(\vec{p}_2)Q_{jk}{}^{f_3}(\vec{p}_2) \bar W_{f_3f_3}(\vec{p}_2)\\
	+2 Q_{f_3}{}^{ik}(-\vec{p}_2)Q_{jk}{}^{f_2}(-\vec{p}_2) \bar W_{f_2f_2}(\vec{p}_2)+\mathcal{O}(\vec{p}_1)\Bigg],
\end{multline}
which holds for all the combinations of fields listed in table \ref{table:combinations unitary}. This is the counterpart of the consistency relation from spatial diffeomorphisms derived in \cite{Berezhiani:2013ewa}. It is a consequence of invariance under spatial diffeomorphisms, analyticity and translation and rotational invariance. The consistency relation simplifies significantly in the soft scalar case $f_1=H_L$, in which it takes the form
\begin{equation}
\frac{\bar{W}_{H_L f_2f_3}(\eta,\vec{p}_1;\eta,\vec{p}_2;\eta)}{\bar{W}_{H_L H_L}(\eta,\vec{p}_1;\eta)}=
	\frac{\delta_{f_2f_3}}{(2\pi)^{3/2}}\left(1-\vec{p}_2\cdot  \frac{\partial}{\partial \vec p_2}\right)\bar{W}_{f_2f_2}(\eta,\vec{p}_2;\eta)
	+\mathcal{O}(\vec{p}_1).
\end{equation}
Note that the right hand side is proportional to $\delta_{f_2f_3}$ because  the three-point function in the soft-momentum limit $H_L(\vec{p}_1\to 0)$ is expected to be roughly the two-point function of the two hard fields $f_2$, $f_3$, which is diagonal in field space.  For two hard scalars, it is also relatively easy to obtain the $\mathcal{O}(\vec{p}_1)$ correction by solving equation (\ref{eq:ST mixed vertex}) to next-to-leading order. Again, one can show that the breaking term can be discarded, thereby yielding \begin{multline}
 \frac{\bar{W}_{H_L H_L H_L}(\eta,\vec{p}_1;\eta,\vec{p}_2;\eta)}{\bar{W}_{H_L H_L}(\eta,\vec{p}_1;\eta)}=
	\frac{1}{(2\pi)^{3/2}}\Bigg[1-\vec{p}_2\cdot  \frac{\partial}{\partial \vec p_2}\\
	+\vec{p}_1\cdot  \frac{\partial}{\partial \vec p_2}
	+(\vec{p}_2\cdot \frac{\partial}{\partial \vec p_2})(\vec{p}_1\cdot  \frac{\partial}{\partial \vec p_2}) 
	-\frac{1}{2} \vec{p}_1\cdot \vec{p}_2 \left( \frac{\partial}{\partial \vec p_2}\right)^2\Bigg]\bar{W}_{H_L H_L}(\eta,\vec{p}_2;\eta)
	+\mathcal{O}(\vec{p}_1{}^2).
\end{multline}
\end{subequations}

\paragraph{Spatially Flat Gauge}
\label{sec:spatial diffs flat}
 Most, if not all, of the consistency relations that have been derived so far apply only in unitary gauge. Yet consistency relations can be also formulated in other gauges. Consider for instance spatially flat gauge, which we recover with a breaking term of the form
\begin{equation}\label{eq:flat gauge}
S_{\mathrm{gf}}= - \frac{M^2}{2}\int d\eta\,  d^3p \left(H_L^2+H_T^2+H_+^2+H_-^2\right)
\end{equation}
when we take $M$ to infinity at the end of the calculation. In this limit, the fields $H_L$, $H_T$ and $H_{\pm}$ decouple from the rest of the perturbations, which amounts to working in a gauge where $H_L\equiv H_T\equiv H_\pm\equiv 0$. In this gauge, the fluctuations of the scalar field $\varphi$ contain all the information about the scalar sector of the perturbations. 

In order to arrive at a consistency relation in spatially flat gauge that follows from spatial diffeomorphisms, we  choose $\alpha=\xi^k(\eta_1,\vec{p}_1)$, $a=f_2(\eta_2,\vec{p}_2)$ and $b=f_3(\eta_3,\vec{p}_3)$ in equation (\ref{eq:consistency start}). Since we have the limit of spatially flat gauge in mind, the fields $f_2$ and $f_3$ therefore stand for either the scalar $\varphi$ or the two tensors $H_{\pm\pm}$. Proceeding as above  we are led to
\begin{multline}\label{eq:ST spatial diffs flat}
	2 a_1^2 p^1_i\bar\Gamma^{ik}{}_{f_2 f_3} \delta_{kj} -
	\Bigg[ 
	\Big(2 p_k^1 Q_{f_2}{}^{ik}(\vec{p}_2) Q_{ij}^{f_3}(\vec{p}_2-\vec{p}_1)
	+(p^2_j-p^1_j) Q_{f_2}{}^{ik}(\vec{p}_2)Q_{ik}{}^{f_3}(\vec{p_2}-\vec{p}_1)\Big)\bar W_{f_3 f_3}(\eta,\vec{p}_2-\vec{p}_1)\\
	+\Big(2 p_k^1 Q_{f_3}{}^{ik}(\vec{p}_1-\vec{p}_2) Q_{ij}^{f_3}(-\vec{p}_2)
	-p^2_j  Q_{f_3}{}^{ik}(\vec{p}_1-\vec{p}_2)Q_{ik}{}^{f_3}(-\vec{p}_2)\Big)\bar W_{f_2 f_2}(\eta,-\vec{p}_2)\\
	+(p^2_j-p^1_j) \delta_{f_2}{}^\varphi \delta_{f_3}{}^\varphi \bar W_{\varphi\varphi}(\eta,\vec{p}_2-\vec{p}_1)
	-p^2_j  \delta_{f_3}{}^\varphi \delta_{f_2}{}^\varphi  \bar W_{\varphi\varphi}(\eta,-\vec{p}_2)
	\Bigg]\frac{\delta(\eta-\eta_1)}{(2\pi)^{3/2}}
=-i(\Gamma_{\Delta S,j})_{f_2f_3},
 \end{multline}
which has essentially the same structure as equation (\ref{eq:ST mixed vertex}), since both capture invariance under spatial diffeomorphisms in the reduced formulation of the theory.  Because the propagator $W_{g\varphi}$ must fall like $1/M^2$, and because $\mathcal{T}$ does not mix metric perturbations and $\varphi$, inspection of equation (\ref{eq:Gamma,alpha m n}) reveals that $(\Gamma_{\Delta S,j})_{f_2f_3}$ vanishes at all orders in $\vec{p}_1$ for the combination of fields listed on table  \ref{table:combinations flat}.   Again, using analyticity we can solve  equation (\ref{eq:ST spatial diffs flat}) by taking partial derivatives wrt $p^1_i$ on both sides of the equation. The unique solution at zeroth order is again given by equation (\ref{eq:K0 mn pq})  with $\partial(\bar{\Gamma}_{\Delta S,j})_{f_2f_3}/\partial p_1^i$ set to zero.  Therefore, the ensuing consistency relations then take the form of equation (\ref{eq:spatial diffs consistency}), where this time $f_1, f_2$ and $f_3$  are drawn from the values listed in table \ref{table:combinations flat}. In this gauge we can for instance reliably determine the cubic vertex $\bar\Gamma^{H_{\pm\pm}}{}_{\varphi\varphi}$ for a soft tensor and two hard scalars. 

\begin{table}  
\begin{center}
\begin{tabular}{c | c | c}
$f_1$ & $f_2$ & $f_3$\\ \hline
$H_{\pm\pm}$ & $\varphi$ & $\varphi$ \\
$H_{\pm\pm}$ & $H_{\pm\pm}$ & $H_{\pm\pm}$\\
\end{tabular}
\end{center}
\caption{The different combination of fields for which the consistency relation (\ref{eq:spatial diffs consistency}) in spatially flat gauge holds. The field $f_1$ carries the soft momentum, whereas the two hard fields $f_2$ and $f_3$ can be interchanged.
\label{table:combinations flat}} 
\end{table}

\subsubsection{Time Diffeomorphisms}

Proceeding along similar lines it is possible to derive a consistency relation that follows from invariance under time diffeomorphisms, among a few other assumptions. To do so, we need to work in spatially flat gauge, as the gauge-fixing condition $\varphi=0$ we used in unitary gauge breaks time diffeomorphisms in an uncontrolled way. One may be  tempted to introduce a symmetry breaking  mass term for the scalar $\varphi$ instead, but this choice is not useful, because the scalar field $\varphi$  would appear in the cubic vertices that are constrained by the Slavnov-Taylor identities (see equation (\ref{eq:0 S})). Because of that, we rather choose  to work in the analogue of spatially flat gauge, with the gauge-fixing terms in equation (\ref{eq:flat gauge}).

We begin the analysis with equation (\ref{eq:consistency start}) for time diffeomorphisms, $\alpha=\xi^0(\eta_1,\vec{p}_1)$, by setting as usual $a=f_2(\eta_2,\vec{p}_2)$ and $b=f_3(\eta_3,\vec{p}_3)$. Factoring out the momentum-conserving delta function we obtain
\begin{multline}\label{eq:ST time diffs flat}
	2a_1^2 \mathcal{H}_1 \delta_{ij} \bar{\Gamma}^{ij}{}_{f_2f_3}(\eta_1,\vec{p}_1;\eta,\vec{p}_2;\eta)+\frac{\partial\bar{\phi}}{\partial \eta_1}\bar{\Gamma}^\varphi{}_{f_2 f_3}(\eta_1,\vec{p}_1;\eta,\vec{p}_2;\eta) \\
	-\frac{\delta(\eta-\eta_1)}{(2\pi)^{3/2}}
	\Bigg\{
	\Big[Q_{f_2}{}^{ij}(\vec{p}_2) Q_{ij}{}^{f_3}(\vec{p}_2-\vec{p}_1)\Big(2\mathcal{H}+\frac{\partial}{\partial\eta}\Big)+\delta_{f_2}{}^\varphi \delta^{f_3}{}^\varphi
	\frac{\partial}{\partial \eta} \Big] \bar{W}_{f_3 f_3}(\eta,\vec{p}_2-\vec{p}_1;\eta)\\
	+\Big[Q_{f_3}{}^{ij}(\vec{p}_1-\vec{p}_2) Q_{ij}{}^{f_2}(-\vec{p}_2)\Big(2\mathcal{H}+\frac{\partial}{\partial\eta}\Big)+\delta^{f_2}{}^\varphi \delta_{f_3}{}^\varphi
	\frac{\partial}{\partial \eta} \Big] \bar{W}_{f_2 f_2}(\eta,-\vec{p}_2;\eta)\Bigg\}=(\bar\Gamma_{\Delta S,0})_{f_2 f_3},
\end{multline}
where the time derivatives only act on the first time argument of the power spectrum. Two different sources can potentially contribute to the symmetry breaking term on the right hand side of the equation: The first is due to the gauge-fixing term (\ref{eq:flat gauge}), but, as in the case of spatial diffeomorphisms in spatially flat gauge,  combining equation (\ref{eq:Gamma,alpha m n}) with equation (\ref{eq:T 0}) we immediately find that in the limit $M\to\infty$ this contribution vanishes for the combination of fields listed on table \ref{table:combinations flat}. The second contribution arises because the reduced formulation of the theory is only invariant under space-independent diffeomorphisms ($\vec{p}_1=0$), but not for time diffeomorphisms with arbitrary spatial dependence. This implies that $(\bar\Gamma_{\Delta S,0})_{f_2 f_3}$ must vanish at zeroth order in $\vec{p}_1$, but not at higher orders. If the two hard fields are of the same type however, $f_2=f_3$,  the invariance of  $(\bar\Gamma_{\Delta S,0})_{f_2 f_2}(\eta_1,\vec{p}_1;\eta_2,\vec{p_2};\eta_3)$ under $\vec{p}_2\to\vec{p}_1-\vec{p}_2$ implies that  the breaking term cannot contain a linear piece in $\vec{p}_1$ either.  In what follows we shall restrict our consideration to the two cases in which the breaking term certainly vanishes. 
 
The problem with equation (\ref{eq:ST time diffs flat}) is that it contains the vertex $\delta_{ij} \bar{\Gamma}^{ij}{}_{f_2 f_3}$, which does not play any role in the limit in which  $H_L$ becomes infinitely heavy. But fortunately, we have already calculated this vertex in the analysis of Section \ref{sec:spatial diffs flat} that led to the solution (\ref{eq:K0 mn pq}), with  $\partial(\bar{\Gamma}_{\Delta S,j})_{f_2f_3}/\partial p^1_i=0$ for the cases listed on table \ref{table:combinations flat}. Substituting that solution into (\ref{eq:ST time diffs flat}) we then obtain a Slavnov-Taylor identity for time diffeomorphisms that contains the relevant vertex $\bar{\Gamma}^\varphi{}_{f_2 f_3}$ alone. Convolving the resulting equation with  the $\varphi$ propagator, integrating  and evaluating at equal times we finally arrive at the consistency relation
\begin{subequations}\label{eq:time consistency relations}
\begin{multline}
\label{eq:time diffs consistency}
	\frac{\bar{W}_{\varphi f_2 f_3}(\eta,\vec{p}_1;\eta,\vec{p}_2;\eta)}{\bar{W}_{\varphi\varphi}(\eta, \vec{p}_1;\eta)}=
	\frac{1}{(2\pi)^{3/2}\bar\phi'}\Bigg\{
	2\Big[\delta_{f_2 f_3}\Big(2\mathcal{H}+\frac{\partial}{\partial\eta}\Big)
	-2\mathcal{H}\delta_{f_2 \varphi} \delta_{f_3 \varphi}
	 \Big] \bar{W}_{f_2 f_2}(\eta,\vec{p}_2)\\
	-\mathcal{H}\Big[
	- p_2^i \frac{\partial \bar W_{f_2 f_2}(\eta,\vec{p}_2)}{\partial p_2^i} \delta_{f_2 f_3}
	+\delta_{f_2 f_3}W_{f_2f_2}(\eta,\vec{p}_2)
	-4 \delta_{f_2 \varphi}\delta_{f_3 \varphi}\bar{W}_{\varphi\varphi}(\eta,-\vec{p}_2)
	\Big]\Bigg\}+\mathcal{O}(\vec{p}_1). 
\end{multline}
This   consistency relation relies on invariance under time diffeomorphisms, although it also depends on the invariance under spatial diffeomorphisms, analyticity and rotational and translational invariance of the quantum state of the perturbations.  Whereas most if not all of the consistency relations that have been discussed in the literature so far involve spatial derivatives of the power spectrum, this relation also contains its time derivatives.

Perhaps because equation (\ref{eq:time diffs consistency}) is quite general, it is a rather formidable expression. We can simplify its form by setting for instance $f_2=\varphi$ and $f_3=\varphi$. This was one of the cases in which the gauge fixing terms do not contribute at zeroth or linear order in momentum, which allows to extend the previous consistency relation to the next order in the soft momentum. The calculation progresses in the same way, the only difference being that the equations for the vertices need to be solved to a higher order.   
\begin{multline}
	\frac{\bar{W}_{\varphi\varphi\varphi}(\eta,\vec{p}_1;\eta,\vec{p}_2;\eta)}{\bar{W}_{\varphi\varphi}(\eta, \vec{p}_1;\eta)}=
	\frac{1}{(2\pi)^{3/2}\bar\phi'}\Bigg[
	2 \frac{\partial \bar W_{\varphi\varphi}(\eta,\vec{p}_2;\eta)}{\partial \eta}-
	\vec{p}_1\cdot\frac{\partial^2 \bar W_{\varphi\varphi}(\eta,\vec{p}_2;\eta)}{\partial \vec{p}_2 \partial \eta}\\
	-\mathcal{H}
	\bigg[-3+(\vec{p}_1-\vec{p}_2)\cdot \frac{\partial}{\partial \vec p_2}	+\Big(\vec{p}_2\cdot \frac{\partial}{\partial \vec p_2}\Big) \Big(\vec{p}_1\cdot  \frac{\partial}{\partial \vec p_2}\Big)-\frac{\vec{p}_1\cdot \vec{p}_2}{2}  \left(\frac{\partial}{\partial \vec p_2}\right)^2\bigg]
	\bar{W}_{\varphi\varphi}(\eta,\vec{p}_2;\eta)\Bigg]
	+\mathcal{O}(\vec{p}_1{}^2). 
\end{multline}
\end{subequations}

\section{Summary and Conclusions}
\label{sec:Summary and Conclusions}

We have explored the constraints that diffeomorphism invariance imposes on the correlation functions of cosmological perturbations. Because these basically follow from symmetry, we have relied on the Lagrangian formulation of the theory, and the corresponding functional integral approach for its perturbative quantization.  In this approach, expectation values can be calculated by introducing a closed time contour. Other than that, the formalism is formally identical to the one  used to calculate in-out matrix elements. 

Our most general constraints take the form of master identities for the generator of connected correlators $iW$ and the generator of one-particle-irreducible diagrams $i\Gamma$ in an  arbitrary gauge. The former are closely related to Schwinger-Dyson equations, which merely state that the classical equations of motion hold in the quantum theory, whereas the latter assume the form of Slavnov-Taylor identities that mirror the (broken) symmetry of the underlying theory. We showed that both sets of identities are equivalent. 

Because diffeomorphism invariance has to be broken in order to quantize the theory, the change of the action under the broken diffeomorphisms plays a crucial role in the Schwinger-Dyson and Slavnov-Taylor identities. The broken symmetry enters through an additional generator, containing an insertion of a single vertex determined by the change of the action under a diffeomorphism transformation. Therefore, these identities are  also a direct reflection of how the symmetry is broken, and not just of the invariance of the theory.  In order to keep such breaking under control, we cannot  gauge-fix   some of the cosmological perturbations to zero, but have to give some of these perturbations a mass term. This is analogous to the use of $R_\xi$ gauges in gauge theories.  As a result, all the fields survive in the gauge-fixed theory, even though some of them are just gauge artifacts in the original  invariant theory.  A compromise emerges if one  lets the symmetry-breaking masses approach infinity, which effectively decouples the corresponding fields from the theory, while keeping the symmetry breaking under control.  In theories in which the metric components $h_{00}$ and $h_{0i}$ are auxiliary fields, it is also possible to  integrate the latter out; the thus ``reduced theory" remains invariant under spatial diffeomorphisms, although it loses invariance under spatially dependent time diffeomorphisms, at least in their original form.

We have formulated our identities in deWitt notation, which allowed us to focus on the conceptual aspects of the identities, rather than on the specific details of diffeomorphism transformations. Consequently, our identities in fact hold in any theory invariant under  a set of symmetries that acts linearly (though possibly inhomogeneously) on the fields. For all those theories, for instance, the Slavnov-Taylor identities state how a three-point function with an insertion of the change of one of the fields under the inhomogeneous component of the  transformation is related to the change of the two-point function solely under the linear component of the transformation.  These identities  provide useful  checks of the self-consistency of the theory, and could be used to diagnose inconsistencies in any calculation of expectation values of cosmological perturbations.

Yet perhaps the most important application of these identities  is the formulation of consistency relations that relate expectation values of products of different numbers of cosmological perturbation fields.  To do so, we had to appeal to  the reduced formulation of the theory in order to limit the total number of  scalars to a single field. In this case, diffeomorphism invariance alone does not suffice to extract definite predictions from the theory, mostly because it is not possible to reduce the scalar field sector to a single field without keeping control of the symmetry.  On the other hand, the additional assumption of analyticity  allowed us to derive consistency relations that constrain the single dynamically relevant field in the gauge-fixed theory, in the limit in which one of the field momenta approaches zero.   We were thus able to reproduce in a different field parameterization the  consistency relations presented in \cite{Berezhiani:2013ewa}. These are captured in our equations (\ref{eq:spatial diffs relations}), which embody consistency relations that follow from spatial diffeomorphisms in unitary gauge, and hold for the set of fields listed on table \ref{table:combinations unitary}.  We  also extended these results to novel relations in spatially flat gauge, still in the reduced formulation of the theory. In this gauge, the consistency relations also take the form of equation (\ref{eq:spatial diffs consistency}), with the fields for which it applies being listed in table \ref{table:combinations flat}.  In the same gauge, we finally derived new consistency relations that follow from invariance under time diffeomorphisms. These relations are listed in equations (\ref{eq:time consistency relations}), and, as opposed to those involving invariance under spatial diffeomorphisms, contain time derivatives of power spectra.

All the consistency relations that we have discussed here are close analogues of the relation between the gravitational vertex and matter self-energy that ultimately enforces the equivalence principle in general relativity. In that sense, unfortunately, our conclusions do not appear to resolve the  tension between Kretschmann's objection to Einstein's principle of general covariance, and the apparent physical  implications of local symmetries, such as the equivalence principle or the consistency relations we just derived. If it turned out that the primordial perturbations did not obey the consistency relations we presented here, we would  probably argue that they were not generated during a period of single-field inflation,  rather than concluding that diffeomorphism invariance is somehow broken.

\section*{Acknowledgments}

C.A.P. and J.T.N. would like to thank Richard Woodard for useful communications and Lasha Berezhiani for stimulating conversations, while R.P. is grateful to Justin Khoury and especially Bart Horn for illuminating discussions. Moreover, R.P. acknowledges the hospitality of the Physics Department at Syracuse University during the initial stages of this collaboration. 
The work of R.P.~was supported by NASA under contract NNX10AH14G and by the DOE under contract DE-FG02-11ER41743.

\appendix

\section{Irreducible Tensors}
\label{sec:Irreducible Tensors}

As we discuss in the main text, in cosmological perturbation theory it is convenient to work with perturbations that transform irreducibly under the isometries of the cosmological background: spatial rotations and translations. We thus introduce a set of eleven irreducible tensors $Q_{\mu\nu}{}^f(\vec{x};\vec{p})$ and $Q^\varphi(\vec{x},\vec{p})$ that we use as basis elements in an expansion of arbitrary cosmological perturbations,
\begin{equation}
	h_{\mu\nu}(\eta,\vec{x})=\sum_{f} \int d^3 p \, Q_{\mu\nu}{}^f (\vec{x};\vec{p})
	\,  f(\eta,\vec{p}), \quad
	\varphi(\eta,\vec{x})=\int d^3 p\,  Q^\varphi(\vec{x};\vec{p})\varphi(\eta,\vec{p}).
\end{equation}
These tensors are plane waves, and although they depend on time through the scale factor,  we suppress the time argument for simplicity,
\begin{equation}
Q_{\mu\nu}{}^f(\vec{x};\vec{p})\equiv a^2 \frac{e^{i\vec{p}\cdot \vec{x}}}{(2\pi)^{3/2}} Q_{\mu\nu}{}^f(\vec{p}), \quad
Q^\varphi(\vec{x};\vec{p})\equiv  \frac{e^{i\vec{p}\cdot \vec{x}}}{(2\pi)^{3/2}}Q^\varphi(\vec{p}),
\end{equation}
with non-vanishing momentum-dependent components 
\begin{subequations}
\begin{align}
{\bf scalars} \quad	 
	Q^\varphi &=1,\\
	Q_{00}{}^A &=-2,\\
	Q_{0i}{}^B &=\frac{i p_i}{p},\\
	Q_{ij}{}^{H_L} &=2 \delta_{ij},\\
		Q_{ij}{}^{H_T} &=2 \left(\frac{1}{3}\delta_{ij}-\frac{p_i p_j}{p^2}\right),
\end{align} 
\end{subequations}
\begin{subequations}
\begin{align}
{\bf vectors} 	\quad Q_{0i}{}^{B_{\pm}} &=- \hat{\epsilon}^\pm_i\\
	Q_{ij}{}^{H_{\pm}} &=-i\left(\frac{p _i}{p}\hat{\epsilon}^\pm_j+\frac{p_j}{p}\hat{\epsilon}^\pm_i\right),
\end{align}
\end{subequations}
\begin{align}
{\bf tensors} 	\quad Q_{ij}{}^{H_{\pm\pm}} &= 2\hat{\epsilon}^\pm_i\hat{\epsilon}^\pm_j.
\end{align}
Here, $\hat{\epsilon}^\pm (\vec{p})$ are two orthonormal transverse vectors with\footnote{These vectors can be taken to be $\hat{\epsilon}^\pm=R(\hat{p})\frac{1}{\sqrt{2}}(\hat{e}_x\pm i \hat{e}_y)$, where $R(\hat{p})$ is a standard rotation mapping the $z$ axis to the $\hat{p}$ direction.}
\begin{subequations}
\begin{align}
	&\vec{p}\cdot \hat{\epsilon}^\pm=0,\\
	 &\vec{p} \times \epsilon^\pm =\mp \, i \,  p \,  \hat{\epsilon}^\pm.
\end{align}
\end{subequations}
Note that the polarization vectors are complex, and that $(\hat{\epsilon}^\pm)^*=\hat{\epsilon}^\mp$. Hence, it follows that $(\hat{\epsilon}^\pm)^*\cdot  \hat{\epsilon}^\pm=\hat{\epsilon}^\mp\cdot  \hat{\epsilon}^\pm=1$, but $\hat{\epsilon}^\pm\cdot  \hat{\epsilon}^\pm=(\hat{\epsilon}^\mp)^* \cdot \hat{\epsilon}^\pm=0$.

Given arbitrary metric and scalar perturbations $h_{\mu\nu}(x)$ and $\varphi(x)$ we would like to find their components in the basis of tensors above. We thus introduce a corresponding set of  projection operators to project onto those components,
\begin{equation}
	f(\eta,\vec{p})=\int d^3 x \, Q_f{}^{\mu\nu}(\vec{p};\vec{x}) h_{\mu\nu}(\eta,\vec{x}),
	\quad
	\varphi(\eta,\vec{p})=\int d^3 p\,  Q_\varphi(\vec{p};\vec{x})\varphi(\eta,\vec{x}).
\end{equation}
These tensors are 
\begin{equation}
	Q_f{}^{\mu\nu}(\vec{p};\vec{x})\equiv \frac{1}{a^2}\frac{e^{-i\vec{p}\cdot \vec{x}}}{(2\pi)^{3/2}} Q_f{}^{\mu\nu}(\vec{p}),\quad
	Q_\varphi(\vec{p};\vec{x})\equiv \frac{e^{-i\vec{p}\cdot \vec{x}}}{(2\pi)^{3/2}} Q_\varphi(\vec{p}),
\end{equation}
where the non-vanishing momentum-dependent components 
\begin{subequations}
\begin{align}
	Q_\varphi &=1,\\
	Q_A{}^{00}{} &=-\frac{1}{2},\\
	Q_B{}^{0i} &=-\frac{i}{2} \frac{p^i}{p}, \\
	Q_{H_L}{}^{ij} &=\frac{1}{6} 
		\delta^{ij},\\
	Q_{H_T}{}^{ij} &=\frac{3}{4}
		\left(\frac{1}{3}\delta^{ij}-\frac{p^i p^j}{p^2}\right),
\end{align}
\end{subequations}
\begin{subequations}
\begin{align}
	Q_{B_{\pm}}{}^{0i} &=-\frac{1}{2} \hat{\epsilon}_\mp^i,\\
	Q_{H_{\pm}}{}^{ij} &=\frac{i}{2} \left(\frac{p^i}{p}\hat{\epsilon}_\mp^j+\frac{p^j}{p}\hat{\epsilon}_\mp^i\right),
\end{align}
\end{subequations}
\begin{align}
	Q_{H_{\pm\pm}}{}^{ij} &=\frac{1}{2}\hat{\epsilon}_\mp^i\hat{\epsilon}_\mp^j,
\end{align}
where vector and tensor indices are raised with the Euclidean metric $\delta^{ij}$ (the location of the indices labeling the projectors $Q$ is a different matter.) These projection operators satisfy the completeness relation
\begin{equation}\label{eq:completeness}
	\int d^3x\,  Q_{f_1}{}^{\mu\nu}(\vec{p}_1;\vec{x}) Q_{\mu\nu}{}^{f_2}(\vec{x};\vec{p}_2)=\delta_{f_1}{}^{f_2} \,\delta^{(3)}(\vec{p}_1-\vec{p}_2)  
\end{equation}
which allows us to project the corresponding fields out a given metric or scalar field fluctuation.
By definition, then, these fields transform as scalars, vectors or tensors. 

It is also convenient to work with the irreducible components of the four-vectors $\xi^\mu$ that parameterize the different infinitesimal diffeomorphisms. We hence write
\begin{equation}
\xi^\alpha(\eta,\vec{x})=\int  d^3x \,
	Q^\alpha{}_{\bar{\alpha}}(\vec{x};\vec{p})\, \xi^{\bar\alpha}(\eta,\vec{p})
	\quad \text{and} \quad
	\xi^{\bar{\alpha}}(\eta,\vec{p})=\int  d^3p \,
	Q^{\bar{\alpha}}{}_\alpha(\vec{p};\vec{x}) \, \xi^\alpha(\eta,\vec{x}),\quad
\end{equation}
where the non-vanishing components of these tensors are
\begin{align}
	Q^0{}_0(\vec{p};\vec{x})&=
	\frac{e^{i\vec{p}\cdot \vec{x}}}{(2\pi)^{3/2}}, 
	\quad
	Q^0{}_0(\vec{x};\vec{p})=
	\frac{e^{-i\vec{p}\cdot \vec{x}}}{(2\pi)^{3/2}},
	\\
	Q^i{}_L(\vec{p};\vec{x})&=-\frac{i p_i}{p}\frac{e^{i\vec{p}\cdot \vec{x}}}{(2\pi)^{3/2}}, \quad
	Q^L{}_i(\vec{x};\vec{p})=\frac{i p_i}{p}\frac{e^{-i\vec{p}\cdot \vec{x}}}{(2\pi)^{3/2}},\\
	Q^i{}_\pm(\vec{p};\vec{x})&=\epsilon^i_{\pm}(\vec{p})\frac{e^{i\vec{p}\cdot \vec{x}}}{(2\pi)^{3/2}}, \quad
	Q^\pm{}_i(\vec{x};\vec{p})=\epsilon_i^{\mp}(\vec{p})\frac{e^{-i\vec{p}\cdot \vec{x}}}{(2\pi)^{3/2}}.
\end{align}

As we discuss in the main text, relations that involve all these projection tensors simplify considerably in deWitt notation. 

\section{Transformation under Diffeomorphisms}
\label{sec:Transformation under Diffeomorphisms}

In order to calculate how the irreducible perturbations introduced above transform under diffeomorphisms, we need to combine equations (\ref{eq:S real}) and (\ref{eq:T real}) with (\ref{eq:Delta transform}). Using the results of Appendix \ref{sec:Irreducible Tensors} we find for time diffeomorphisms ($\alpha=\xi^0$),  longitudinal diffeomorphisms ($\alpha=L$) and transverse diffeomorphisms of either helicity ($\alpha=\pm$)
\begin{subequations}\label{eq:S fields}
\begin{equation}\label{eq:0 S}
\mathcal{S}_{f(\eta_1,\vec{p}_1)}{}_{\xi^0(\eta_2,\vec{p}_2)}=
\delta(\vec{p}_1-\vec{p}_2)\left[\delta_f{}^A \left(\mathcal{H}_1+\frac{d}{d\eta_1}\right)-p_1\delta_f{}^B+\mathcal{H}_1 \delta_f{}^{H_L} +\frac{\partial \bar{\phi}}{\partial  \eta_1}\delta_f{}^\varphi \right]\delta(\eta_1-\eta_2)
\end{equation}
\begin{equation}\label{eq:longitudinal S}
	\mathcal{S}_{f(\eta_1,\vec{p}_1)}{}_{\xi^L(\eta_2,\vec{p}_2)}=
	\delta(\vec{p}_1-\vec{p}_2)
	\left[-\delta_f{}^B \frac{d}{d\eta_1}
	+\frac{p_1}{3}\delta_f{}^{H_L}
	- p_1 \delta_f{}^{H_T} \right]\delta(\eta_1-\eta_2),
\end{equation}
\begin{equation}\label{eq:transverse S}
	\mathcal{S}_{f(\eta_1,\vec{p}_1)}{}_{\xi^\pm(\eta_2,\vec{p}_2)}=
	\delta(\vec{p}_1-\vec{p}_2)
	\left[-\delta_f{}^{B_{\pm}} \frac{d}{d\eta_1}
	-p_1 \delta_f{}^{H_{\pm}}\right]\delta(\eta_1-\eta_2).
\end{equation}
\end{subequations}
In these equations a prime denotes a derivative with respect to conformal time $\eta$, $\mathcal{H}\equiv a'/a$ and $p\equiv |\vec{p}|$. The  components of the transformations linear in the fields are
\begin{subequations}
\begin{multline}\label{eq:T 0}
	\mathcal{T}_{f_1(\eta_1,\vec{p}_1)}{}^{f_2(\eta_2,\vec{p}_2)}{}_{\xi^0(\eta_3,\vec{p}_3)}=
	\frac{\delta(\vec{p}_2+\vec{p}_3-\vec{p}_1)}{(2\pi)^{3/2}}\Bigg[\delta_{f_1}^A \delta^{f_2}_A \left(2\mathcal{H}_1-\frac{\partial}{\partial \eta_2}-2\frac{\partial}{\partial \eta_3}\right)
	\\
	-4i p^3_i Q_{f_1}{}^{0i}(\vec{p}_1) \delta^{f_2}_A 
	{}+Q_{f_1}{}^{0i}(\vec{p_1})Q_{0i}{}^{f_2}(\vec{p}_2)\left(4\mathcal{H}_1-2\frac{\partial}{\partial \eta_2}-2\frac{\partial}{\partial \eta_3}\right)
	+2i p^3_j Q_{f_1}{}^{ij}(\vec{p}_1) Q_{i0}{}^{f_2}(\vec{p}_2) +\\
	{}+Q_{f_1}{}^{ij}(\vec{p}_1)Q_{ij}{}^{f_2}(\vec{p}_2)\left(2\mathcal{H}_1-\frac{\partial}{\partial \eta_2}\right)
	-\delta_{f_1}^\varphi \delta_\varphi^{f_2}
	\frac{\partial}{\partial \eta_2}\Bigg]\delta(\eta_1-\eta_2)\delta(\eta_1-\eta_3),
\end{multline} 
\begin{multline}\label{eq:T k}
\mathcal{T}_{f_1(\eta_1,\vec{p}_1)}{}^{f_2(\eta_2,\vec{p}_2)}{}_{\xi^k(\eta_3,\vec{p}_3)}=
	\frac{\delta(\vec{p}_2+\vec{p}_3-\vec{p}_1)}{(2\pi)^{3/2}}
	\Bigg[
	\delta_{f_1}^A Q_{0k}{}^{f_2}(\vec{p}_2)
	\frac{\partial}{\partial \eta_3}
	+i\delta_{f_1}^A \delta^{f_2}_A p^2_k+\\
	{}+2i Q_{f_1}{}^{0i}(\vec{p}_1) Q_{0k}{}^{f_2}(\vec{p_2}) p^3_i
	-2 Q_{f_1}{}^{0i}(\vec{p}_1)  Q_{ik}{}^{f_2}(\vec{p}_2)
	\frac{\partial}{\partial \eta_3}
	+2i Q_{f_1}{}^{i0}(\vec{p}_1)Q_{i0}{}^{f_2}(\vec{p}_2)
	p^2_k+\\
 {}+2iQ_{f_1}{}^{ij}(\vec{p}_1)Q_{ik}{}^{f_2}(\vec{p}_2) p^3_j
	+i Q_{f_1}{}^{ij}(\vec{p}_1) Q_{ij}{}^{f_2}(\vec{p}_2)
	p^2_k
	+i \delta_{f_1}^\varphi \delta_\varphi^{f_2}p^2_k	\Bigg]
	\delta(\eta_1-\eta_2)\delta (\eta_1-\eta_3), 
\end{multline}
\end{subequations}
where the non-vanishing components of the tensors $Q_{f}{}^{\mu\nu}(\vec{p})$ and $Q_{\mu\nu}{}^f(\vec{p})$ are given in Appendix \ref{sec:Irreducible Tensors}. Contacting equation (\ref{eq:T k}) with $Q^k{}_{L}$ and $Q^k{}_{\pm}$ one readily recovers the transformations under longitudinal and transverse diffeomorphisms.  Note that for some choices of the fields, these expressions can be further simplified. For instance, for $f_1=H_L$,
$Q_{f_1}{}^{ij}(\vec{p}_1) Q_{ij}{}^{f_2}(\vec{p}_2)=\delta_{H_L}^{f_2}$. 

\end{fmffile}

\end{document}